\documentclass[12pt,fleqn]{article}
\usepackage{amsmath,a4,latexsym,epsfig}
\newcommand{\intd}{\int \! d^4 x \;}
\newcommand{\intS}{\int \! d S \;}
\newcommand{\intSbar}{\int \! d \bar S \;}
\newcommand{\intV}{\int \! d V \;}
\newcommand{\Ga} {\Gamma}
\newcommand{\Gacl} {{\Gamma_{\rm cl}}}

\renewcommand{\L}{{\cal L}}
\newcommand{\Abar}{{\bar A}}
\newcommand{\Tr}{{\rm Tr}}
\newcommand{\lambdabar}{{\overline\lambda}}
\newcommand{\sigmabar}{{\overline\sigma}}
\newcommand{\psibar}{{\overline\psi}}
\newcommand{\Fbar}{{\bar F}}
\newcommand{\phibar}{{\overline \varphi}}
\newcommand{\epsilonbar}{{\overline\epsilon}}
\newcommand{\thetabar}{{\overline\theta}}
\newcommand{\etabar}{{\overline\eta}}
\newcommand{\chibar}{{\overline\chi}}
\newcommand{\qbar}{{\overline q}}
\newcommand{\fbar}{{\overline f}}
\newcommand{\cbar}{{\overline c}}

\newcommand{\alphadot}{{\dot\alpha}}

\newcommand{\lambdaV}{{\tilde \lambda}}
\newcommand{\lambdaVbar}{{\overline {\tilde \lambda}}}
\newcommand{\DV}{{\tilde D}}

\def\dF#1{\frac{\delta{\cal F}}{\delta#1}}

\def\df#1{\frac{\delta}{\delta#1}}

\def\pslash#1{{\setbox0=\hbox{$#1$}
  \rlap{\ifdim\wd0>.7em\kern.22\wd0\else\kern.1\wd0\fi /}#1}}

\def\brs{\mathbf s}
\newcommand{\mn}{\mu\nu}

\begin{document}
\begin{titlepage}

\begin{flushright}
MPI-PhT/2001-24\\
BN--TH--01--06\\
{\tt hep-th/0107239}\\
\end{flushright}

\begin{center}
{\large\bf{An anomalous breaking of supersymmetry  in \\[1ex]
        supersymmetric 
  gauge theories with local  coupling}}
\\
\vspace{8ex}
{\large       Elisabeth Kraus}
{\renewcommand{\thefootnote}{\fnsymbol{footnote}}
\footnote{E-mail address:
                kraus@th.physik.uni-bonn.de,\\
Presently at {\it Max-Planck-Institut f\"ur Physik (Werner-Heisenberg-Institut) \\
\indent\indent F\"ohringer Ring 6, 80805 Munich, Germany}}} 
\\
\vspace{4ex}
{\small\em                Physikalisches Institut,
              Universit{\"a}t Bonn,\\
              Nu{\ss}allee 12, D--53115 Bonn, Germany\\
}
\vspace{2ex}
\end{center}
\vfill
{\small
 {\bf Abstract}
 \newline\newline
 We 
extend the gauge coupling of Super--Yang--Mills theories to an
external real superfield composed out of a chiral and an antichiral
superfield
and perform renormalization in the extended model.
In one-loop order
we find an anomalous breaking of
supersymmetry which vanishes in the limit to constant coupling.
 The anomaly
arises from non--local contributions and  its
coefficient  is gauge  and scheme independent and strictly of one-loop
order. In the perturbative
framework of the construction
the anomaly cannot be absorbed as a counterterm to the classical
action. 
With local gauge coupling the symmetric counterterms with chiral
integration are holomorphic functions and this property is
independent from the specific
regularization scheme.
 Thus, the symmetric counterterm to the
Yang--Mills 
part is of one--loop order only  -- and it is the anomaly which gives rise to the
two--loop order of the gauge $\beta$--function in pure
Super--Yang--Mills theories and to its closed all-order expression.

\vfill
}

\end{titlepage}


\newpage
\section{Introduction}

The extension of coupling constants to space-time dependent external
fields, i.e.~to local couplings, has been an important tool in
renormalized perturbation theory for a long time \cite{Bogo,EPGL}. It has
been moved in the center of interest again  from a string point of
view since the couplings there are dynamical fields  and
 enter quite naturally as external fields the effective field theories
derived from 
string theories. 
Inspired from string theory  local
couplings in combination with  holomorphicity have been
considered also in quantum field theory  \cite{SHVA91,Louis91,SEI93,Louis94}.

Also in ordinary quantum field theory
 local
couplings found recently interesting applications:
In the Wess--Zumino model \cite{FLKR00}, in SQED \cite{KRST01} and
 in softly broken SQED \cite{KRST01soft} they allowed
the derivation of the
non-renormalization theorems \cite{FULA75,GSR79}
from  renormalization properties of the
extended model in an algebraic context. 

The same analysis is here applied to Super--Yang--Mills theories: We
extend the gauge coupling to an external real superfield by
introducing a chiral and an antichiral field into the Yang--Mills part.
Since these fields compose the real superfield of the coupling, 
they are not independent fields
 and their imaginary part couples to a total
derivative in the action. 
This property can be expressed by a Ward identity.

By means of this Ward identity  symmetric counterterms to chiral vertices are
holomorphic functions of the chiral field composing the supercoupling.
 In particular
counterterms to the Yang-Mills part are
exhausted in one-loop order and this property just expresses
 the generalized non-renormalization theorem of the coupling \cite{SHVA86}.

However, if the coupling is treated as a space-time dependent field,
supersymmetry has an anomalous breaking in one-loop order.
 The anomalous breaking of supersymmetry is
different from the Adler--Bardeen anomaly \cite{AD69} in its algebraic
characterization: It is the variation of a local field monomial, but
the corresponding field monomial  involves the logarithm of the gauge
coupling.  The perturbative expansion and in particular all
loop diagrams are power series in the gauge coupling. Thus,
the anomaly is determined by  non-local and finite parts  of 
Green functions, which are not subject of renormalization.
 This property characterizes the anomalous breaking of
supersymmetry as a true anomaly. 

Indeed,  we are able to prove by algebraic consistency that the 
 coefficient  of the anomaly is scheme  and gauge
independent.  Hence, it appears 
 in the manifest supersymmetric gauge and in the
Wess--Zumino gauge irrespective of the specific scheme used for the
subtraction of divergences. However we point out, that it might be
shifted by non-invariant counterterms
from  supersymmetry  to the constraining Ward identity of the
 chiral and antichiral fields composing the gauge coupling. And
with a supersymmetric invariant
scheme for the subtraction of divergences as the BPHZ scheme
 the anomaly will appear  
 as an anomalous breaking of the latter identity.

As a specific and very important application 
 we work out the implications of the anomaly for
the gauge $\beta$-function in the Wess--Zumino gauge.
 There, the
 anomaly
naturally appears 
as a breaking of
supersymmetry in the Slavnov--Taylor identity. 
We show that the anomaly can be absorbed into the 
Slavnov--Taylor operator  by modifying the  supersymmetry
transformations of the local gauge coupling. When we include the
matter part a second modification appears due to the Adler--Bardeen
anomaly of the axial current \cite{KRST01}. Then algebraic
 renormalization can be performed on the basis of the anomalous
 Slavnov--Taylor operator. The Callan--Symanzik equation is restricted
 by the symmetries of the theory, and we find, that
 algebraic consistency  with the
constraining Ward identity of the chiral and antichiral fields restricts
the $\beta$-function of the local gauge coupling to one-loop order and that
algebraic consistency with the 
the anomaly part of
 Slavnov--Taylor identity
determines the two-loop  term of the gauge $\beta$-function in
terms of the anomaly coefficients and the anomalous mass dimension.
Hence we find a closed expression of the gauge $\beta$-function in
 Super--Yang--Mills theories by passing over from a constant coupling to an
 external superfield.

The plan of the paper is as follows: In section 2 we introduce the
classical action of supersymmetric non-abelian gauge theories with the
local gauge coupling and an axial vector multiplet. Section 3 is
devoted to the supersymmetry anomaly: We give the defining symmetry
relations and prove its scheme  and gauge independence. In section 4 we
construct the anomalous Slavnov--Taylor identity in the Wess--Zumino
gauge and perform algebraic renormalization including the axial vector
multiplet and the Adler--Bardeen anomaly in section 5. Finally as
an application we derive the closed expression of the gauge $\beta$
function from an algebraic construction of
the Callan--Symanzik equation and the renormalization
group equation. In the conclusions we discuss how our results
elucidate previous ones, in particular those of \cite{SHVA86,SHVA91, ARMU97},
and how the notion of holomorphic dependence \cite{Louis91,SEI93}
should be properly
understood, if one takes local couplings seriously in quantum field theory.
In appendix A we list the BRS transformations of fields, in appendix B 
 the complete symmetric operators of the Callan--Symanzik equation and
 renormalization group equation. Conventions and notations are the
ones given  in appendix A of \cite{KRST01}

\newpage

 \section{The classical action} 

We consider supersymmetric non-abelian
gauge theories with a simple gauge group as for example $SU(N)$ 
including   chiral and antichiral
matter 
multiplets  $A^ k_L, \Abar^k_L$ and $A^k_R, \Abar^k_R$,
\begin{equation}
\label{chiralmult}
A^ k = \varphi^k  + \sqrt 2  \theta^ \alpha \psi^k_{\alpha} + F^ k \theta^2 
 \  ,\quad\hbox{and}\quad
\Abar^ k = \phibar^k +   \sqrt 2\psibar^k_{\alphadot} \thetabar^ \alphadot +
 \Fbar^ k   \thetabar^2
\end{equation}
which  transform under an irreducible representation of the gauge group:
\begin{eqnarray}
\delta_{\omega(x)}^ {\rm gauge}A^ k_L & = & i \omega_a (x)(T_a)^ k{ }_{j} A^
j_L 
\ ,\qquad \delta_{\omega(x)}^ {\rm gauge}\Abar_ {Lk} = - i \omega_a (x)
\Abar_ {Lj} 
(T_a)^ j{ }_{k}\ , \nonumber \\
\delta_{\omega(x)}^ {\rm gauge}\Abar^ k_{R} & = & i \omega_a (x)(T_a)^ k{
}_{j} \Abar ^
j_R \ ,
\qquad \delta_{\omega(x)}^ {\rm gauge}A_ {Rk} = - i \omega_a (x)
 A_ {Rj} 
(T_a)^ j{ }_{k} \ ,
\end{eqnarray}
and
\begin{equation}
\big[ T_a , T_b \bigr] = i f_{abc} T_c \ .
\end{equation}
We impose parity conservation and invariance under
charge conjugation throughout  the paper.

 In the Wess--Zumino gauge  the
vector multiplet consists of the gauge fields, the gauginos and the
auxiliary fields $D$:
\begin{eqnarray}
\label{Amult}
\phi^ A &= & \theta \sigma^ \mu \thetabar A^\mu - i \thetabar^2  \theta^ \alpha \lambda _\alpha +i \theta^2 \thetabar_\alphadot\lambdabar^ \alphadot 
+ \frac 14 \theta^2 \thetabar^2  D \ , \\
\phi^ A &= & \phi^ A_a \tau_a  \nonumber \qquad \mbox{with} \quad \Tr (\tau_a \tau_b) = 1\ .
\end{eqnarray}
Here $\tau_a$ are the matrices of the fundamental representation.

The classical action composed of these fields
is invariant under non-abelian gauge
transformations and the non-linear supersymmetry transformations of the
Wess--Zumino gauge \cite{WitFree}

In a recent paper \cite{KRST01}
the non-renormalization theorems of SQED and its implications for the
renormalization group functions have been
derived      
by extending the gauge
coupling to an external real superfield and by taking into account
softly broken axial symmetry with its anomaly. The classical construction is
based on the supersymmetry transformations of the gauge invariant 
Lagrangians and can be used for the non-abelian  gauge theories
without modifications. In analogy to SQED we introduce 
in addition to the physical fields the following external field multiplets: 
\begin{itemize}
\item  We extend the gauge coupling $g$ to an external field multiplet
 $G (x, \theta, \thetabar )$, whose lowest component is the local
gauge coupling $g(x)$. 
From the chiral and antichiral nature of the Yang--Mills Lagrangian it
 is seen 
that the supercoupling   $G (x, \theta, \thetabar )$  is
a constrained real superfield
 being composed of a chiral and
 an
antichiral
field multiplet $\mbox{\boldmath{$\eta$}}(x, \theta)$ and
$\mbox{\boldmath{$\etabar$}} (x, \thetabar)$:
\begin{eqnarray}
\label{E2def}
G (x, \theta, \thetabar ) & = & ({\mbox{\boldmath{$\eta$}}}(x, \theta,
\thetabar ) +{\mbox{\boldmath{$\etabar$}}}(x, \theta, \thetabar ))^ 
{-\frac 12} \equiv g(x) + {\cal O}(\theta, \thetabar)
\end{eqnarray} 
with 
\begin{equation}
\label{defeta}
{\mbox{\boldmath{$\eta$}}}(x, \theta) = \eta + \theta^ \alpha
 \chi_\alpha + \theta^2 f \ ,
 \qquad 
{{\mbox{\boldmath{$\etabar$}}} }(x, \thetabar) = \etabar + \theta_\alphadot \chibar ^
 \alphadot  + \thetabar^2 {\overline f} 
\end{equation}
in the chiral and antichiral representation, respectively.
\item  Supersymmetric gauge theories with a local gauge supercoupling
are complete in the sense of multiplicative renormalization only when
we include an axial multiplet (see   eq.~(\ref{ctmatter}) and
cf.~also the discussion in~\cite{KRST01}): 
\begin{equation} 
\label{Vdef}
\phi^V =   \theta \sigma ^ \mu \thetabar V^ {\mu} -i \thetabar^2 \theta^ \alpha
 \lambdaV_ \alpha + i \theta^2 \thetabar_\alphadot \lambdaVbar^
 \alphadot + \frac 14  \theta^2 \thetabar ^2 \DV .
\end{equation}
The  vector multiplet transforms as a singlet under the gauge group.
The interaction of the axial vector field with the matter fields is
governed by $U(1)$ axial symmetry.
\begin{eqnarray}
\delta_{\tilde \omega}^{\rm axial} A_L^ k = - i \tilde \omega (x) A_L^ k \ ,
\qquad
\delta_{\tilde \omega}^{\rm axial} \Abar_L^ k = i \tilde \omega (x)
\Abar_L^ k \ ,\\ 
\delta_{\tilde \omega}^{\rm axial} A_R^ k = - i \tilde \omega (x) A_R^ k\ ,
\qquad
\delta_{\tilde \omega}^{\rm axial} \Abar_R^ k =  i \tilde \omega (x)
\Abar_R^ k \ .
\nonumber 
\end{eqnarray}
\item
At the classical level axial symmetry is broken softly by the matter
 mass terms. For this reason we introduce a
  chiral and and an antichiral field multiplet $\mathbf q$ and $
\mathbf \qbar$ with dimension one:
\begin{equation}
\label{qdef}
{\mathbf q} =
q + \theta q^ \alpha + \theta^2 q_F \quad \mbox{and} \quad
{\mathbf \qbar} =
 \qbar + \thetabar _\alphadot\qbar^ \alphadot + \thetabar^ 2 \qbar_F.
\end{equation} 
They transform with respect to gauge symmetry as singlets  and
with respect to axial symmetry with charge two  including a shift in the
lowest component:
\begin{equation}
\delta_{\tilde \omega}^{\rm axial} {\mathbf q} =  2 i \tilde \omega 
({\mathbf q} + m) \ .
\end{equation}
When we couple the $\mathbf q$-multiplets
 to the matter mass term, softly broken
axial symmetry can be expressed as a Ward identity for the classical action.
\end{itemize}

The invariant classical action consists of the 
 Yang-Mills part $\Ga_{\rm YM}$, the matter part
$\Ga_{\rm matter}$ and the matter mass term $\Ga_{\rm mass}$ enlarged by
 the chiral and antichiral interaction of the $\mathbf q$-multiplets
 $\Ga_q$:
\begin{equation}
\label{Gasusy}
{\Ga_{\rm susy}} = \Ga_{\rm YM} + \Ga_{\rm matter} + (\Ga_{\rm mass} + \Ga_q)
\end{equation}
The individual parts
are  invariant under non-abelian gauge transformations with the
local gauge coupling and $U(1)$ axial symmetry and they are
supersymmetric. Using the superfield expressions for the field multiplets
 it is possible to
express the invariant action 
as a superspace integral
\begin{eqnarray}
\label{GaYM}
\Ga_{\rm YM} &= &- \frac 14\intS {\mbox{\boldmath{$\eta$}}}  
{\cal L}_{\rm YM}
- \frac 14 \intSbar {{\mbox{\boldmath{$\etabar$}}}}  \bar {\cal L}_{\rm YM} \\
\label{Gamatter}
\Ga_{\rm matter} & = & \frac 1 {16}\intV \L_{\rm matter} \\
\label{Gamass}
\Ga_{\rm mass} + \Ga_q &= & 
- \frac 1 4\intS ({\mathbf q} + m ) \L_{\rm mass}
- \frac 1 4\intSbar ({\mathbf \qbar} + m ) \bar \L_{\rm mass}
\end{eqnarray}
and
\begin{eqnarray}
\label{LYM}
\L_{\rm YM} &\equiv & \Tr W^ \alpha W_{\alpha } = W^ \alpha_a W_{\alpha a } \ ,  \quad
W_\alpha \equiv  \frac 1 {8\sqrt 2 } \bar D \bar D (e^ {-2g(x) \phi^ A } D_\alpha 
e^ {2g(x) \phi^ A })\ ; \\  
\label{Lmatter}
\L_{\rm matter} & = &  \Abar_{Lk} e^ {- 2 g \phi^ A_a (T_a)^k{
}_j + 2 \phi^ V \delta^ k { }_j} A^ j_L +
A_{Rk} e^ { 2 g \phi^ A_a (T_a)^k{
}_j + 2 \phi^ V \delta^ k{ }_j} \Abar^ j_R  \ ,\\
\label{Lmass}
\L_{\rm mass}  &= & 
 A_L^k A_{R k}\ .
\end{eqnarray}
For vanishing axial vectors and
in  the limit to constant supercoupling $G(x, \theta, \thetabar) \to
g=$const.\ the action (\ref{Gasusy})
 is the usual action of Super--Yang--Mills theories with matter:
\begin{equation}
\lim_{G\to g} \Ga_{\rm susy}\Big|_{\phi^V = 0}  = \Ga_{\rm SYM}
\end{equation}

In the classical action, only
the Yang--Mills part depends  on the supercoupling
via      the chiral and antichiral
multiplets
${\mbox{\boldmath{$\eta$}}}$ and ${\mbox{\boldmath{$\etabar$}}}$.
The dependence of   the matter part on the local
gauge coupling $g(x)$  stems  from a
redefinition of the vector multiplet,
\begin{equation}
\phi^ A \to g(x) \phi^A \ ,
\end{equation}
compared to the usual normalization.
  In perturbation theory such a  redefinition is important
for obtaining a  bilinear free field action
 and associated free field propagators.\footnote{ Alternatively, we
  could 
proceed in the conventional normalization but with 
  a shift in the lowest component of the chiral multiplets. The
  implications of both constructions are equivalent.}

In the present construction the chiral and antichiral fields
${\mbox{\boldmath{$\eta$}}}$ and ${\mbox{\boldmath{$\etabar$}}}$ are
not independent fields, but compose the real superfield of the
supercoupling (see (\ref{E2def})).
 Therefore the imaginary part of their lowest components
appears only with a total derivative:
\begin{equation}
\Bigl(\df{ \eta } - \df {\etabar} \Bigr)\Ga_{\rm susy} = - \frac i 4 \epsilon^{\mu
\nu \rho \sigma }G_{\mn a} G_{\rho \sigma a} + i \partial _\mu(\lambda_a
\sigma^ \mu \lambdabar_a ) \ .
\end{equation}
This property gives rise to the following identity
\begin{equation}
\label{holomorphsusy}
\intd\Bigl(\df{ \eta } - \df {\etabar} \Bigr)\Ga_{\rm susy} = 0 \ .
\end{equation}
In effect, this equation restricts the  symmetric chiral counterterms to
holomorphic functions of ${\mbox{\boldmath{$\eta$}}}$ and
${\mbox{\boldmath{$\etabar$}}}$ and has important implications for the
renormalization properties of chiral Green functions and of the
coupling.  

For proceeding to higher orders of perturbation theory the invariant classical
action has to be supplemented by the gauge fixing of gauge vector
fields:
\begin{equation}
\Ga_{\rm g.f} = \intd \Bigl(\frac 12 (\xi(x) + \xi)B_a B_a + B_a \partial_\mu
A^ \mu_a + H^ \mu B_a A_{\mu a}\Bigr)
\end{equation}
We have introduced a local gauge parameter $\xi(x)$ and an additional
vector field $H^  
\mu$ for later use in the Callan--Symanzik equation.

The gauge fixing
 breaks non-abelian gauge invariance and supersymmetry. For this
reason we have to combine the symmetry transformations of the fields
into  BRS-transformations \cite{White92a,MPW96a}.
 Introducing the Faddeev-Popov fields $c_a$,
 the axial ghost $\tilde c$ and the supersymmetry and translational
 ghosts $\epsilon^ \alpha $,
$\epsilonbar^ \alphadot$ and $\omega^ \mu$
the BRS-operator acts on the
fields of the gauge invariant action (\ref{Gasusy})
as
\begin{equation}
\label{BRS}
\brs \phi = (\delta^ {\rm gauge}_{c(x)} + \delta^ {\rm axial}_{\tilde
c(x)} + \epsilon ^ \alpha \delta_\alpha + \bar \delta_\alphadot
\epsilonbar^ \alphadot - i \omega^ \mu \partial_\mu) \phi \ .
\end{equation}
 BRS-transformations of the ghosts are determined by the structure
constants of the algebra and the algebra of symmetry transformations
is expressed in  nilpotency of the BRS operator. 
 The BRS transformations of the
fields are summarized in  appendix A. 

By means of  BRS transformations the gauge fixing can be extended by
the ghost part to a BRS invariant action:
\begin{equation}
\label{Gagf}
\Ga_{\rm g.f.} + \Ga_{\rm ghost} = \brs \intd \bigl( \frac 12
\xi \bar c B + \bar c  \partial^\mu A_\mu  + H^ \mu \bar c_a
 A_{\mu a}\bigr)
\end{equation}
The explicit form can be worked out with the BRS transformations of
the appendix:
The ghost part contains the usual Faddeev-Popov action of non-abelian
gauge theories and in addition compensating terms for supersymmetry
depending on the supersymmetry ghosts $\epsilon_\alpha$ and
$\epsilonbar_\alphadot$ \cite{White92a,MPW96a,HKS99}. In addition,
we have introduced  a BRS partner $\chi_\xi$ to the gauge fixing
parameter and a BRS partner $C^ \mu$ to the field $H^ \mu$.
By means of the BRS varying gauge parameter it is possible
to
identify gauge parameter independent quantities by their algebraic
characterization as  non-BRS variations \cite{PISI85,HAKR97}.

Adding an external field part $\Ga_{\rm ext}$ for the non-linear
BRS transformations of propagating fields
(see (\ref{extf})) one can express BRS invariance of the classical 
action and the
algebra of symmetry transformations by the
Slavnov--Taylor identity in the usual way. At this stage we finally 
eliminate the auxiliary fields $D$ (\ref{Amult}) and $F_L $ and $F_R$
and their complex conjugates
(\ref{chiralmult}) by their
equations of motion.
The complete classical action
\begin{equation}
\Ga_{\rm cl}= \Ga_{\rm susy}
 + \Ga_{\rm g.f. } + \Ga_{\rm ghost} + \Ga_{\rm ext}
\end{equation}
satisfies the Slavnov--Taylor identity 
\begin{equation}
\label{STclassical}
{\cal S} (\Gacl) = 0
\end{equation}
with the usual Slavnov--Taylor operator (see (\ref{STOperator}))
and it satisfies the identity (\ref{holomorphsusy})
\begin{equation}
\label{holomorphcl}
\intd\Bigl(\df{ \eta } - \df \etabar \Bigr)\Gacl = 0 \ .
\end{equation}
 
On the basis of the classical action loop calculations are immediately
performed by treating the local coupling and its superpartners as
external fields. Green functions with the local coupling are
determined by differentiation with respect to the local coupling and
performing the limit to constant coupling:
\begin{equation}
\label{Greendef}
\Ga_{g ^ n \varphi_{i_1}\cdots \varphi_{i_m} } (x_1, \cdots x_n, y_1, \cdots y_m)  = \lim_{G \to g}
\frac {\delta^{n+m} \Ga } {\delta g (x_1) \cdots \delta g (x_n) \delta
\varphi_{i_1} \cdots \varphi_{i_m}}\Big|_{\varphi_i = 0}
\end{equation}
Here the fields $\varphi_i$ summarize propagating and external fields of the
theory.

The local gauge coupling $g(x)$ is distinguished from ordinary external
fields by the 
property that it is the perturbative expansion parameter. 
 For any 1PI Green function (\ref{Greendef}) the power of the
constant gauge coupling is determined by the loop order $l$,
by the number of amputated external
legs $N_{\rm amp. legs} $ and by the number of external field differentiations.
This property 
 can be expressed by the topological formula:
\begin{equation}
\label{topfor}
N_{g(x)} = N_{\rm amp. legs} + N_Y+ 2N_f+ 
2N_\chi + 2 N_{\eta - \etabar} 
+ 2(l-1)\ .
\end{equation}
Here $N_Y$ denotes the number of BRS insertions,
and $N_f$, $N_\chi$ and $N_{\eta - \etabar } $ gives the number of 
insertions 
corresponding to  the respective external fields and their complex
conjugates. The topological formula is valid for the classical action
with $l= 0$.
 
For constant coupling $g(x) \to g$ 
the 1PI Green functions are power series in the
coupling and result for constant supercoupling $G \to g$
 in the 1PI Green functions of ordinary
supersymmetric gauge theories:
\begin{equation}
\lim_{G\to g} \Ga\Big|_{\phi^ V = 0\atop \tilde c = 0} =  \Ga^ {\rm SYM}\ .
\end{equation}
Thus,  renormalization properties derived from the extended theory
are satisfied also for the ordinary Green functions of supersymmetric
non-abelian gauge theories.

\section{The anomaly in Super--Yang--Mills theories with local gauge coupling}
\subsection{Characterization of the anomalous breaking}

For the purpose of the present section we set the fields of the axial vector
multiplet (\ref{Vdef})
 and  the axial ghost $\tilde c$ to zero and consider the 
renormalization of
supersymmetric non-abelian gauge theories with the local gauge
supercoupling
(\ref{E2def}).
In the procedure of renormalization the possible
breakings of the defining symmetries have to be classified according
to their properties as scheme dependent breaking terms and scheme
independent breaking terms, the anomalies. Scheme dependent breakings
are introduced in the procedure of renormalization via the subtraction
of ultraviolet divergences and can be removed by adjusting suitable
counterterms to the classical action. Anomalies arise from non-local
contributions, which are not subject of renormalization, and cannot be
absorbed into counterterms.

For the Adler--Bardeen  anomaly, the distinguishing mark to 
scheme dependent breakings  is its algebraic property as 
being not a gauge variation of a four-dimensional field monomial.
For the supersymmetry anomaly found in the present paper 
 the
situation is different:
From the algebra of supersymmetry transformations it is quite
 generally deduced that the breakings of supersymmetry are variations
\cite{PSS80}. 
However, with local gauge coupling we find a breaking of supersymmetry 
which can be written as the supersymmetry transformation of a field
monomial involving the logarithm of the coupling.

Indeed when we work out the 
 the BRS variation of the gauge 
 invariant term
\begin{equation}
\label{lng}
\intd \ln g(x) (L_{\rm YM} + \bar L_{\rm YM})\ ,
\end{equation}
we find a breaking of supersymmetry, which depends on the logarithm of
the coupling only via a total derivative:
\begin{eqnarray}
\label{Deltaanomalybrs}
\Delta^{\rm anomaly}_{\rm brs} & =   & \brs \intd \ln g(x) (L_{\rm YM} + \bar
L_{\rm YM}) \\
& = & (\epsilon^ \alpha \delta_\alpha + 
\epsilonbar^ \alphadot \delta_\alphadot) 
\intd \ln g(x) (L_{\rm YM} + \bar
L_{\rm YM}) \nonumber\\
& = & \intd  \Bigl( \ln g(x) i \bigl(\partial _\mu \Lambda_{\rm YM} ^
\alpha \sigma ^ \mu_{\alpha \alphadot} \epsilonbar^ \alphadot -
\epsilon^ \alpha \sigma^ \mu_{\alpha \alphadot}  \partial_\mu \Lambda
^ {\alphadot}_{\rm YM} \bigr) \nonumber \\
& & \phantom{\intd} - \frac 12 g^ 2 (x) (\epsilon \chi + \chibar
\epsilonbar)
(L_{\rm YM} + \bar L _{\rm YM}) \Bigr)\nonumber\ .
\end{eqnarray}
Here
 $L_{\rm
YM}$ and $\Lambda^\alpha$ are the   $F$ and spinor component
of the chiral multiplet  $\L_{\rm YM}$ (\ref{LYM}):
\begin{equation}
\L_{\rm YM} = - \frac 12 \lambda_a \lambda_a + \theta ^\alpha \Lambda
_\alpha + \theta^2 L_{\rm YM}\ .
\end{equation}
 Hence,  $\Delta_{\rm brs}^{\rm anomaly}$ is independent of $\ln g$
 in the limit to  constant coupling and for any differentiation
 with respect to the coupling (see (\ref{Greendef})).

 Since the perturbative expansion is a
 power series expansion in the coupling, the local field monomial
 (\ref{lng}) is not related to ultraviolet divergent diagrams.
Thus the corresponding breaking
 results from non-local contributions, which are not subject of
 renormalization. 
For this reason it has to be considered 
as an anomaly of supersymmetry appearing in supersymmetric gauge theories
with local gauge coupling.

Similarly, applying  the identity (\ref{holomorphcl})
to the gauge invariant 
and supersymmetric  expression
\begin{equation}
 \frac 14 \intS \ln {\mbox{\boldmath{$\eta$}}}\, \L_{\rm YM} + 
\frac 14 \intSbar \ln
{{\mbox{\boldmath{$\etabar$}}}}\, {\overline
\L}_{\rm YM} \ ,
\end{equation}
we find again an expression which is independent of a logarithm and
 can appear as a breaking of the identity (\ref{holomorphcl}):
\begin{eqnarray}
\label{Deltaanomalyeta}
\Delta^ {\rm anomaly}
_{\eta-\etabar} & = & \frac 14 
\intd \Bigl(\df {\eta(x)} - \df {\etabar(x)}\Bigr)
     \bigl(
\intS \ln {\mbox{\boldmath{$\eta$}}}\, \L_{\rm YM} +  \intSbar \ln
{{\mbox{\boldmath{$\etabar$}}}}\, {\overline
\L}_{\rm YM} \bigr) \nonumber  \\
& =  &  \frac 14 \Bigl(
\intS  {\mbox{\boldmath{$\eta$}}}^ {-1}\, \L_{\rm YM} + \intSbar 
{{\mbox{\boldmath{$\etabar$}}}}^ {-1}\, {\overline
\L}_{\rm YM}\Bigr)\ . 
\end{eqnarray}

The field monomials  $\Delta^{\rm anomaly}_{\rm brs}$ and
 $\Delta^{\rm anomaly}_{\eta-
\etabar}$ satisfy the topological formula in one loop-order and 
could  appear both
as breakings of the respective symmetry identities.
However, 
 we will show in the next section, that it is possible to add ordinary 
counterterms which are power series in the coupling  in
such a way, that either the Slavnov--Taylor identity or the
$\eta-\etabar$ identity is fulfilled in its classical form
remaining with an anomalous breaking in the complementary identity,
i.e.\  either
\begin{eqnarray}
\label{STDeltabrs}
& & \intd \bigl(\df{\eta} - \df{\etabar}\bigr) \Ga = 0 \ , \nonumber\\
& & {\cal S} (\Ga) = r_{\eta}^ {(1)} \Delta^{\rm anomaly}_{\rm brs} + {\cal O}(\hbar^ 2)\ ;
\end{eqnarray}
or
\begin{eqnarray}
\label{etaDeltabrs}
& & \intd \bigl(\df{\eta} - \df{\etabar}\bigr) \Ga = \frac 12 r_{\eta}^
{(1)}
\Delta^{\rm anomaly} _{\eta- \etabar} + {\cal O} (\hbar^2 )\ , \nonumber\\
& & {\cal S}(\Ga) = 0\ .
\end{eqnarray}
can be established.

The coefficient  $r_{\eta}^ {(1)}$ is shown to be the 
 gauge and scheme independent. For the classical Lagrangian specified
 in (\ref{Gasusy})
 $r_{\eta}^ {(1)}$ is determined independently from the matter
 interaction
 and takes the value
\begin{equation}
\label{reta}
r_{\eta}^ {(1)} =  \frac {C(G)} {8 \pi^2 }  \ .
\end{equation}
For SQED  it is zero as  can be seen 
from eqs.~(76) and (78) in ref.~\cite{KRST01}.

In the Wess--Zumino gauge the
calculation of the anomaly coefficient 
is quite involved and requires to solve various equations
of the Slavnov--Taylor identity for the unique definition of the
anomalous field monomial.  Details of the calculation are presented
elsewhere and we restrict ourselves for 
the remaining part of the paper to work out its implications for the
$\beta$-functions of Super--Yang--Mills theories. From this construction the
coefficient is implicitly determined by the two-loop gauge
$\beta$-function.

\subsection{The algebraic construction}

For a systematic construction of the  anomalous breaking and its
defining symmetries we use the quantum action principle and algebraic
consistency. Therefrom we find that the breaking of the classical
symmetries (\ref{STclassical}) and (\ref{holomorphcl})
 is  local 
\begin{eqnarray} \label{STbreaking}
& & {\cal S} (\Ga) = \Delta_{\rm brs} + {\cal O} (\hbar^2 )\ , \nonumber\\
\label{etabreaking}
& &\intd (\df {\eta} - \df{\etabar} ) \Ga =  \Delta_{\eta - \etabar} +
{\cal O} (\hbar^2 ) \ ,
\end{eqnarray}
and restricted by the following consistency equations:
\begin{eqnarray}
\label{Deltabrscons}
\brs_\Gacl \Delta_{\rm brs}  &= &0 \ ,
\nonumber \\
\brs_\Gacl \Delta_{\eta- \etabar}& = & \intd \bigl(\df {\eta} - \df
{\etabar}\bigr) \Delta_{\rm brs} \ .
\end{eqnarray}
The power of  the gauge coupling  in the local field
monomials
$\Delta_{\rm brs}$ and $\Delta_{\eta - \etabar} $ is determined by the
topological formula (\ref{topfor}).

In a first step we consider the field monomials contributing to 
 $\Delta_{\eta-\etabar}$.
As long as one does not 
 care about the Slavnov--Taylor identity
 they  can be
trivially absorbed by adjusting local counterterms as power series
in $\eta - \etabar$ and we get: 
\begin{equation}
\label{holomorphqu}
\intd \bigl(\df {\eta} - \df {\etabar}\bigr) \Ga = 0 \ .
\end{equation}
 In the
BPHZL scheme with ordinary $M(s-1)$ mass terms for massless fields and
in dimensional regularization with MS-subtraction the
identity (\ref{holomorphqu}) is fulfilled by construction since the
insertions defined by $\eta- \etabar$  are total derivatives.
Having established (\ref{holomorphqu}),  the consistency equation
(\ref{Deltabrscons}) implies that 
$\Delta_{\rm brs}$  depends on $\eta - \etabar $ only by a total
derivative.

Considering now the field monomials in $\Delta_{\rm brs}$
it is seen that the anomalous breaking
  $\Delta^ {\rm anomaly}_{\rm brs}$ (\ref{Deltaanomalybrs}) 
is indeed a possible
breaking of the Slavnov--Taylor identity in one-loop order:
 It is $\brs_\Gacl$-invariant
and it satisfies the topological formula. However, being the variation
of the field monomial (\ref{lng}) it cannot be absorbed into a counterterm.
  $\Delta^{\rm anomaly}_{\rm brs}$ is 
the only anomalous breaking of the Slavnov--Taylor identity.
 The BRS variations of similar terms
\begin{equation}
\label{matterln}
\intd g ^ k \ln g L_{\rm matter} \quad \mbox{and} \quad
 \intd g^ k \ln g (L_{\rm mass} + \bar
L _{\rm mass} )
\end{equation}
 depend on logarithms  for $k \neq 0$. For $k= 0$ they do not satisfy
the topological formula in loop order  $l \geq 1$. 
Hence, the corresponding
loop diagrams are absent and the BRS variations of (\ref{matterln})
 cannot appear as  breakings of
the Slavnov--Taylor identity.

Since the  Adler--Bardeen anomaly of gauge symmetry 
is absent due to parity conservation the breaking
$\Delta^{\rm anomaly}_{\rm brs}$ is the only term
 which cannot be absorbed as a counterterm  and
we have established:
\begin{eqnarray}
\label{ST1loop}
{\cal S} (\Ga) \Big|_{\phi^V = 0\atop \tilde c = 0}
 &= & r^{(1)}_\eta \Delta^ {\rm anomaly} _{\rm brs} + {\cal
O}(\hbar^2) \nonumber \\
\intd \Bigl(\df \eta - \df \etabar\Bigr)\Ga &=& 0 \ .
\end{eqnarray}

Having identified the anomaly in the Slavnov--Taylor identity we can
finally adjust counterterms in such a way that the Slavnov--Taylor
identity is unbroken, and the anomaly is shifted to the
$\eta-\etabar$ identity. Using 
\begin{equation}
\brs_{\Gacl} \Bigl(\intS \ln {\mbox{\boldmath{$\eta$}}}\, \L_{\rm kin} + \intSbar \ln
{{\mbox{\boldmath{$\etabar$}}}}\, {\overline
\L}_{\rm kin} \Bigr)= 0
\end{equation}
and
\begin{equation}
\ln {{\mbox{\boldmath{$\etabar$}}}} = - \ln (2 g^2) + \ln(1  +  g^2 (\eta - \etabar) + 2 \theta
\chi g^2 +   
2 \theta^ 2 f g^2 )
\end{equation}
it is possible to rewrite $\Delta^{\rm anomaly}_{\rm brs}$ into
\begin{equation}
\Delta^{\rm anomaly}_{\rm brs} = 
-\frac 18 \brs _{\Gacl}
 \Bigl(\intS \ln(1 +  g^2 (\eta - \etabar) + 2 \theta \chi g^2 +  
2 \theta^ 2 f g^2) \L_{\rm YM} + \mbox{c.c.}  \Bigr) \ .
 \end{equation}
The counterterm
\begin{equation}
\label{cteta}
\Ga_{{\rm ct}, \eta- \etabar} =
 \frac 18 r_\eta^ {(1)} \intS \bigl(\ln(1 + 2 g^2 (\eta - \etabar) + 2 \theta \chi g^2 +  
 \theta^ 2 f ) \L_{\rm YM} + \mbox{c.c} \bigr)
\end{equation}
is an admissible non-invariant counterterm: It does not involve
logarithms of the coupling but contains an ordinary power series
expansion in external fields. Changing
\begin{equation}
\Ga \to \Ga ' = \Ga + \Ga_{{\rm ct}, \eta- \etabar}
\end{equation}
the anomalous breaking is removed from the Slavnov Taylor identity,
\begin{equation}
{\cal S} (\Ga') = 0 \ .
\end{equation}
but 
 the  identity (\ref{holomorphqu}) is broken in
one-loop order by
$\Delta^ {\rm anomaly}_{\eta-\etabar}$
\begin{equation}
\intd{\Bigl(\df \eta- \df \etabar\Bigr)} \Ga' = 
\intd{\Bigl(\df \eta- \df \etabar\Bigr)} \Ga_{{\rm ct}, \eta- \etabar} = 
\frac 12 r_\eta^ {(1)}
 \Delta^ {\rm anomaly}_{\eta-
\etabar} + {\cal O}(\hbar^ 2)
\end{equation}
Hence,  the  anomaly of supersymmetric
 gauge theories with local gauge coupling
 can be transformed from one
symmetry identity to another, but it cannot be arranged to vanish at
all by adjusting a local counterterm in the action.

From general principles there is no preference for keeping the
 Slavnov--Taylor identity or  the $\eta-\etabar$ identity
 (\ref{holomorphcl})
in their
 classical form. 
Indeed, dependent on the gauge we will prefer one of the both
alternatives:
In the supersymmetric gauge  it is appropriate to shift the
anomaly into the $\eta- \etabar$ identity loosing otherwise the simple
notion of fields in superspace.  In the Wess--Zumino gauge of the
present paper, we take
the anomaly in the Slavnov--Taylor identity and leave the
$\eta-\etabar$ identity unmodified to all orders. As we will show in
section 4,
this
 choice has the
advantage, that the Slavnov--Taylor operator for 
 the Adler--Bardeen anomaly of the $U(1)$ axial symmetry is strictly
of  one-loop order
and takes the same form as in SQED.

\subsection{Scheme  and gauge independence of the anomaly}

Using the BRS varying gauge parameter it is obvious from the
construction that $r_\eta^ {(1)}$ is gauge  independent:
When we  include the
gauge parameter into the field monomial (\ref{lng}) its BRS variation
 depends on the logarithm of the
coupling in combination with the BRS partner of  $\xi$:
\begin{equation}
\brs  \intd  \xi \ln g(x) (L_{\rm YM} +\mbox{c.c.}) =
\xi \Delta_{\rm brs}^ {\rm anomaly} +  \chi_\xi \intd 
\ln g(x) (L_{\rm YM} + \mbox{c.c}) 
\end{equation}
Since diagrams with logarithms of the coupling are not present, we
conclude:
\begin{equation}
\partial_ \xi r^{(1)}_\eta = 0\ .
\end{equation}

Scheme independence of $r^{(1)}_\eta$ can be inferred as usual	 from the 
renormalization group  equation: 
Introducing in one-loop order a scale parameter $\kappa$, which
might be a normalization point  or a scheme dependent scale as
the unit mass $\mu$ in dimensional regularization or even a Wilsonian
cutoff, the classical action is independent of the scale, i.e.
\begin{equation}
\kappa \partial _\kappa \Gacl = 0\ .
\end{equation}
In one-loop order the equation is broken  
by local field monomials:
\begin{equation}
\kappa \partial_\kappa \Gacl = \Delta_\kappa
\end{equation}
Once more, since diagrams
with logarithms of the coupling are absent, 
the field monomials in $\Delta_\kappa$ are polynomials  in the coupling
according to  the topological formula.
Using the consistency equation with the Slavnov--Taylor operator
\begin{equation}
\brs_{\Ga} \kappa\partial_\kappa (\Ga) -  \kappa \partial_\kappa {\cal S}
(\Ga) = 0
\end{equation}
 we
find from (\ref{ST1loop}):
\begin{equation}
\label{kappacons}
\kappa\partial_\kappa r^{(1)}_\eta \Delta_{\rm brs}^ {\rm anomaly} =
\brs_{\Gacl} \Delta_\kappa\ .
\end{equation}
By rewriting $\Delta^{\rm anomaly}_{\rm brs} $ into the $\brs_{\Gacl}$-variation of
the monomial (\ref{lng}), which involves  the logarithm of the
coupling, it is seen that the right- and left-hand-side of
eq.~(\ref{kappacons}) can only match  if 
\begin{equation}
\kappa\partial _\kappa r^{(1)}_\eta=  0\ ,
\end{equation}
i.e.~$r^{(1)}_\eta $ is scheme independent.

As it is seen in the following part of the paper, the one-loop anomaly
implies the two loop order  of the gauge $\beta$-function in pure
Super--Yang--Mills  theories, which would be in
naive applications of classical symmetries or with vanishing $r^{(1)}_\eta$
indeed of  order one-loop
only without matter or SQED-like with matter.

 From the characterization of the anomaly, as a scheme and gauge
independent quantity, it is obvious that the anomaly appears
independent from the procedure of regularization. Consequently an 
an invariant scheme, which preserves gauge invariance and
supersymmetry for Super--Yang--Mills theories with local gauge coupling,   cannot
exist. 
     If one wants   to remove the anomalous breaking 
from  supersymmetry and the $\eta-\etabar$ identity at the same time,
  the  counterterm (\ref{lng})
with the logarithm of the coupling 
has to be
adjusted with the scheme independent coefficient $r_\eta^{(1)}$. The
addition of such a counterterm does not correspond to a redefinition
of a loop subtraction and is in principle excluded from the
perturbative construction.
Doing this nevertheless,
 the perturbative expansion becomes much worse than the
ordinary power series expansion since one starts
a  logarithmic series simultaneously with the ordinary power series in one
loop order. In two-loop order the logarithmic series has the same anomaly as
the perturbative series in one-loop order, starting there a $\ln^2$-series
in perturbation theory. 
All 1PI Green functions in $l\geq 2 $
depend then on the logarithms of
the coupling and  the limit to constant coupling does not result in
ordinary
Super--Yang--Mills  theories but in a quantum field theory defined by a
logarithmic series in the coupling.

\section{The Slavnov--Taylor operator of the anomaly} 

For   proceeding with renormalization to higher orders 
the  anomaly  has to be absorbed in
 form of an operator into the Slavnov--Taylor identity (see
(\ref{STDeltabrs}))
or the $\eta- \etabar$ identity (see (\ref{etaDeltabrs})). 
Both alternatives are possible, for the present paper we use the
version (\ref{STDeltabrs}), where the $\eta-\etabar$ identity is
unbroken. Then 
the anomaly  can be absorbed into the  Slavnov--Taylor operator by
modifying the supersymmetry transformations of the gauge coupling.

From its explicit expression 
(\ref{Deltaanomalybrs})
it is seen that we
can rewrite $\Delta^ {\rm anomaly} _{\rm brs}$ into the following
operator form:
\begin{eqnarray}
\Delta^{\rm anomaly}_{\rm brs} & = &  r^{(1)}_\eta \intd \Bigl(\frac 12 g^ 5 (x) (\epsilon^
\alpha \chi_\alpha + \chibar_\alphadot \epsilonbar^ \alphadot)\df
{g(x)} \\
& &\phantom{r_\eta \intd} - 2i \frac 1 {g(x)} \partial _\mu g(x) 
\bigl((\sigma^ \mu \epsilonbar)^ \alpha \df {\chi^ \alpha} +
(\epsilon \sigma^ \mu) ^ \alphadot \df{\chibar^ \alphadot} \bigr)\Bigr)
\Gacl \nonumber \\
& &  + \frac {r^{(1)}_\eta} 2 \brs_{\Gacl} \intd \Bigl( \frac 12 g^ 4 
(\epsilon^
\alpha \chi_\alpha + \chibar_\alphadot \epsilonbar^ \alphadot) 
\Tr \bigl((\rho^ \mu - \partial^ \mu\bar c) A_\mu - Y_c c) \bigr)\ . \nonumber
\end{eqnarray}
 Redefining $\Ga $ by the  non-invariant counterterm
\begin{equation}
\Ga \to \Ga + \frac {r^{(1)}_\eta} 2 \Tr \intd 
(\epsilon^
\alpha \chi_\alpha + \chibar_\alphadot \epsilonbar^ \alphadot) g^ 4
\bigl((\rho^ \mu - \partial^
\mu\bar c) A_\mu + Y_c c) \bigr)
\end{equation}
the redefined action satisfies the following modified Slavnov--Taylor identity:
\begin{eqnarray}
\label{STmod}
{\cal S }(\Ga) & - &
 r^{(1)}_\eta \intd \Bigl(\frac 12 g^ 5 (x) (\epsilon^
\alpha \chi_\alpha + \chibar_\alphadot \epsilonbar^ \alphadot)\df
{g(x)} \\
& &\qquad - 2i \frac 1 {g(x)} \partial _\mu g(x) 
\bigl((\sigma^ \mu \epsilonbar)^ \alpha \df {\chi^ \alpha} +
(\epsilon \sigma^ \mu) ^ \alphadot \df{\chibar^ \alphadot} \bigr)\Bigr)
\Ga =  0 + {\cal O} (\hbar^2 ) \ .\nonumber
\end{eqnarray}
Apparently,
 the additional one-loop operators are just modifications
of the supersymmetry transformations of the local gauge coupling.

For algebraic consistency nilpotency of the Slavnov--Taylor operator
(\ref{nilpotency})
has to be fulfilled also for the modified Slavnov--Taylor operator
(\ref{STmod}).
In the present case, requiring nilpotency  is
nothing but establishing the supersymmetry algebra on the local
coupling. A straightforward calculation proves that the supersymmetry
algebra closes if we take the following modifications in the
transformation of the gauge coupling and its fermionic superpartner
\begin{eqnarray}
\label{brsanomalous}
 \brs^{r_\eta} g^ 2 &=& - g^ 4 (\epsilon^
\alpha \chi_\alpha + \chibar_\alphadot \epsilonbar^ \alphadot) F(g^
2) - i \omega^\nu\partial_\nu g^2\ , \\
 \brs^{r_\eta} (\eta- \etabar)& = & (\epsilon^
\alpha \chi_\alpha + \chibar_\alphadot \epsilonbar^ \alphadot)
- i \omega^\nu\partial_\nu (\eta - \etabar)\ ,
\nonumber\\
\brs^{r_\eta} \chi^ \alpha & = & 2 \epsilon^ \alpha f + i (\sigma^  \mu
\epsilonbar)^ \alpha
(\partial_\mu g^2 \frac 1 { F (g^2)} + \partial_\mu (\eta - \etabar)
- i \omega^\nu\partial_\nu \chi^\alpha \ , \nonumber \\
\brs^{r_\eta} \chibar^ \alphadot &= & - 2 \epsilonbar^ \alphadot \fbar + i (
\epsilon \sigma^  \mu
)^ \alphadot
(\partial_\mu g^2 \frac 1 {F (g^2)} - \partial_\mu (\eta - \etabar)
- i \omega^\nu\partial_\nu \chibar^\alphadot\ ,  \nonumber
\end{eqnarray}
leaving the BRS transformations of $f$ and $\fbar$ unchanged
(cf. (\ref{BRScoupling})).
The function $F(g^ 2)$ is a power series in the square of the coupling, i.e.
\begin{equation}
F(g^ 2) = 1 + r^{(1)}_\eta g^2 + {\cal O} (g^ 4).
\end{equation}
The transformation (\ref{brsanomalous}) involves an infinite power series
via $1/ F(g^2)$ in the supersymmetry
transformation of the spinors, whenever the one-loop coefficient $r^{(1)}_\eta$ is
non-vanishing.

The one-loop order in the expansion is determined by the anomaly
coefficient and is scheme independent, it cannot be generated by a
redefinition of the coupling.
 Higher orders correspond to a redefinition of the
coupling and are scheme dependent. To prove
this,
we define
a new coupling $\hat g$ by
\begin{equation}
\label{grednonp}
{\hat g}^2  = \frac {g^2} {1+  r^{(1)}_\eta g^ 2 \ln g^ 2 + \sum_{l\geq 2} z^
{l}g^ {2l}}\ .\end{equation}
Evaluating $\brs^{r_\eta} \hat g^2(g^2)$ 
 it is seen
that the anomalous one-loop term is removed in the supersymmetry
transformation of $g^2$. The coefficients $z^{(l)}, l\geq 2$ can be
adjusted 
 in such a way that the supersymmetry
transformation of $\hat g$ takes the original form:
\begin{equation}
\delta_\alpha \hat g ^2 =  - \chi_\alpha \hat g^4\ .
\end{equation}
 As for the counterterms  the redefinition (\ref{grednonp})
with the logarithm of
the coupling is in contradiction to the power series expansion of loop
diagrams. 
Therefore,  the one-loop term $r^{(1)}_{\eta}$ in the modified supersymmetry
transformations (\ref{brsanomalous})
is unambiguously defined by one-loop diagrams  of the
theory, whereas the coefficient of order $l \geq 2$
 correspond to finite redefinitions and
are scheme dependent.

 In general, these redefinitions are
determined by normalization conditions on the gauge coupling. In the
opposite way, 
fixing the higher orders of $F(g^ 2)$ in the Slavnov--Taylor identity
determines the normalization of the coupling. 
Taking, for example,
 the transformations of spinors in (\ref{STmod}) as exact, we have to choose
\begin{equation}
\label{FNSZV}
F(g^ 2) = \frac 1 {1 -r^{(1)}_\eta g^2}\ .
\end{equation}
With this function 
we get finally the NSVZ expression for the gauge $\beta$-function
\cite{NSVZ83}.
Taking the transformation of $g^2$ in (\ref{STmod}) as exact,
\begin{equation}
\label{F2loop}
F(g^ 2) =  {1 +r^{(1)}_\eta g^2} \ ,
\end{equation}
one obtains a gauge $\beta$-function  which is vanishing in
loop order $l \geq 3$ in pure Super--Yang--Mills. 
However, none of these choices is distinguished
from other ones, as long as they are not related to a specific
normalization condition for physical Green functions.

Modifying the BRS transformations of the local coupling $g(x)$
and  its fermionic superpartners $\chi^ \alpha,\chi^ \alphadot$
according to eq.~(\ref{brsanomalous}) an anomalous
 Slavnov--Taylor identity and the $\eta-\etabar$ identity
can be established to all orders:
\begin{equation}
\label{STreta}
{\cal S}^ {r_\eta} (\Ga) = 0 \quad \mbox{and} \quad \intd \Bigl(\df{\eta} -
\df{\etabar} \Bigr)\Ga = 0
\end{equation}
with
\begin{eqnarray}
{\cal S}^ {r_\eta} (\Ga) = {\cal S}(\Ga) & - & \intd \Bigl(g^ 4 \delta F(g^ 2)(\epsilon
\chi + \chibar \epsilonbar) \df {g^2} \\ & & \quad - 
i \frac {\delta F }{1+ \delta F}
 \partial_\mu g^ 2 \bigl (\sigma ^ \mu \epsilonbar)^\alpha
\df {\chi^ \alpha}  + (\epsilon \sigma^ \mu )^ \alphadot \df {\chibar^
 \alphadot} \bigr) \ ,\nonumber 
\end{eqnarray}
and (see (\ref{brsanomalous}))
\begin{equation}
\delta F(g^2) \equiv F(g^2) -1 = r^{(1)}_\eta g^2  + {\cal O}(g^4) \ .
\end{equation}
 Since we have avoided to adjust counterterms with
logarithms of the coupling, the perturbative expansion is an expansion
in powers of the gauge coupling and an anomalous breaking as the one
of the one-loop order is excluded in higher orders by the topological
formula (see (\ref{matterln})).

\section{Renormalization and non-renormalization\\ theorems}

For a complete description of supersymmetric non-abelian gauge
theories with matter we have to complement the defining symmetries of
the model by the $U(1)$ axial  symmetry. At the quantum level axial
symmetry is broken by the Adler--Bardeen anomaly \cite{AD69}, which appears in the
present construction as a second anomalous breaking of the
Slavnov--Taylor identity in one-loop order \cite{KRST01}: 
\begin{equation}
\label{gaugeanomalie}
\Delta_{\rm brs}^ {\rm axial}\Big|_{\epsilon, \epsilonbar, \omega= 0}  =  
r^ {(1)} \Tr \intd\tilde c  \epsilon^{\mu \nu \rho\sigma}
G_{\mn} (gA )
G_{\rho\sigma} (gA) \ .
\end{equation}
The anomaly coefficient $r ^ {(1)}$ is determined
 by the usual triangle diagrams:
\begin{equation} 
\label{r1}
r^ {(1)} = - \frac {1} {16 \pi^ 2} C(G)\ .
\end{equation}

We want to mention that in the present construction
with local gauge coupling the non-renormalization theorem of the
Adler--Bardeen anomaly \cite{ADBA69} is  a simple consequence of global axial
symmetry and is not inferred from additional properties of the loop
expansion.\footnote{For a more detailed discussion see \cite{KRST01}, where
the same arguments have been 
used for the non-renormalization of the axial anomaly in  SQED.}
Taking into account  
the topological formula,  the gauge anomaly of loop order  $l$
takes the  general form:
\begin{equation}
\Delta_{\rm brs}^ {{\rm axial}(l)}\Big|_{\epsilon, \epsilonbar, \omega= 0}  = 
r^ {(l)} \Tr \intd g^ {2(l-1)}(x)\tilde c  \,  \epsilon^{\mu \nu \rho\sigma}
G_{\mn} (gA )
G_{\rho\sigma} (gA) \ .
\label{anomalygeneral}
\end{equation}
Only in one loop order the coefficient of the ghost is a total
derivative, and for this reason the anomalous breaking appears in the
Ward identity of global $U(1)$ axial symmetry for $l\geq 2$. Since the global
symmetry is unbroken, the coefficients of the anomalous breaking
vanish identically in loop order $l\geq 2$
\begin{equation}
r^ {(l)} = 0 \qquad \mbox{for} \quad l \geq 2 \ .
\end{equation}

By means of the  local gauge supercoupling
the Adler--Bardeen anomaly with its supersymmetric extension 
can be expressed by a $\brs_\Ga$-invariant
operator acting on the classical action \cite{KRST01}:
\begin{eqnarray}
\label{operatoranomalie}
\Delta ^ {\rm axial}_{\rm brs}
& = & 4 i r^{(1)} \intd  \Bigl(\tilde c \Bigl(\frac {\delta}{\delta \eta} -
\frac {\delta}{\delta \etabar}\Bigr) + 2i (\epsilon \sigma^ \mu )^
\alphadot V_\mu \frac {\delta}{\delta \chibar^ \alphadot} 
 - 2i ( \sigma^ \mu \epsilonbar )^ \alpha V_\mu \frac
 {\delta}{\delta \chi^ \alpha} \nonumber \\ { } & &
 + 2 \epsilonbar_\alphadot \lambdaVbar ^\alphadot 
\frac {\delta}{\delta f} - 2 \lambdaV^\alpha \epsilon_\alpha \frac {\delta}{\delta
 \fbar} \Bigr) \Ga_{\rm cl} \nonumber \\
& \equiv & - r^ {(1)} \delta {\cal S}\Ga _{\rm cl}\ .
\end{eqnarray}
In this form it can be included into an anomalous
 Slavnov--Taylor operator and we end with the final version of the
 Slavnov--Taylor identity of Super--Yang--Mills theories with matter
 and with local gauge coupling:
\begin{equation}
\label{STanomalous}
{\cal S}^ {r_\eta} (\Ga) + r^{(1)}\delta {\cal S} \Ga = 0.
\end{equation}
Here
 $\delta {\cal S}$ is the operator expressing the Adler--Bardeen
anomaly (\ref{operatoranomalie}) 
and ${\cal S}^ {r_\eta} (\Ga)$ is the operator (\ref{STreta}) which includes
the modified supersymmetry transformations of the gauge coupling.
The anomalous Slavnov--Taylor operator has the same nilpotency properties
 as the classical
Slavnov--Taylor operator (\ref{nilpotency}).

The Slavnov--Taylor identity (\ref{STanomalous}) 
is supplemented by the identity
\begin{equation}
\label{holomorphqu2}
\intd (\df \eta - \df \etabar) \Ga = 0 \ .
\end{equation}
Only if  the Slavnov--Taylor identity is combined with the identity
(\ref{holomorphqu2}),
 the anomaly coefficient $r^{(1)}_ \eta$ and its higher-order
 extensions  and the operator $\delta
{\cal S}$ are unambiguously determined. 

On the basis of the anomalous Slavnov--Taylor identity 
(\ref{STanomalous}) and the identity
(\ref{holomorphqu2})  the renormalization of
supersymmetric gauge theories with local gauge coupling
can be performed in
 algebraic renormalization independent from properties of a specific scheme. 
Obviously, it is not possible to find  
a general classical solution of the anomalous
Slavnov--Taylor identity and, as a consequence,  
 an invariant scheme cannot be constructed for supersymmetric 
non-abelian gauge theories with
local gauge
coupling. 
Thus,
the Slavnov--Taylor identity has to be solved order by order  and established by
adjusting non-invariant counterterms to the classical action, if necessary.

 In each order
there remain as undetermined coefficients the coefficients of
symmetric counterterms, which are invariants of the classical
symmetries:
\begin{equation}
\brs _{\Gacl} \Ga_{\rm ct,inv} = 0\ , \qquad \intd (\df \eta - \df
\etabar) \Ga_{\rm ct, inv} = 0\ .
\end{equation}

It is the remarkable feature of the construction with the local
coupling
that the
gauge independent symmetric  counterterms are  restricted by the 
$\eta-\etabar$ identity
(\ref{holomorphqu2}): For chiral integrals the symmetric counterterms
are indeed holomorphic functions of the chiral fields $\mbox{\boldmath{$\eta$}}$ and
$\mbox{\boldmath{$\etabar$}}$ and symmetric counterterms to chiral
vertices are absent. The only
invariant counterterm with chiral integration which is left is
 the one-loop counterterm to the Yang-Mills part of the classical
action:
\begin{eqnarray}
\label{ctYM}
\Ga^ {(1)}_{\rm ct, YM} &= &\frac 18
\Bigl(\intS \L_{\rm YM} + \intSbar\bar\L_{\rm YM}\Bigr) \ .
\end{eqnarray}
 These restrictions reflect the specific renormalization
properties of supersymmetric gauge theories \cite{KRST01}
summarized in the
non-renormalization theorems in superspace \cite{FULA75,GSR79,SHVA86}.

There appears a second gauge independent $l$-loop counterterm to the
matter part of the action, which takes in superfield notation the
form (cf.~(\ref{Lmatter})):
\begin{eqnarray}
\Ga^ {(l)}_{\rm ct, matter} &= &\frac 1 {16}
\intV G^ {2l}\L_{\rm matter} \label{ctmatter} \ .
\end{eqnarray}
It is this counterterm, which enforced the introduction of
the axial vector multiplet,
since the expression (\ref{ctmatter})  cannot be related to a
field renormalization of matter fields as long as the coupling is local.

Due to the bilinear form of the Slavnov--Taylor identity
 there appear gauge dependent field
redefinitions to the individual fields of the theory, i.e. these are
field redefinitions of the vector fields and its superpartners, field
redefinitions
of the matter fields and a field redefinition of the Faddeev-Popov
ghost  $c$:
\begin{eqnarray}
\label{fieldred}
A \to (1 + \delta z_A) A\ , &\qquad &\lambda \to (1+ \delta z_\lambda) \lambda
\ , \qquad c \to (1 + \delta z_c) c \ ,\\
\psi_A\to (1 + \delta z_\psi)\psi_A \ , &\qquad & \varphi_A \to (1+
\delta z_\varphi) \varphi_A 
\ ,\qquad A= L, R \nonumber 
\end{eqnarray}
With  local coupling we have in addition the following two generalized
field redefinitions of Weyl spinors into scalar superpartners:
\begin{equation}
\label{fieldredgen}
\lambda_\alpha \to \lambda_\alpha + \delta z_{\lambda A}i (\sigma^\mu\chibar)_\alpha
A_\mu\ ,\quad
 \psi_\alpha \to \psi_\alpha + \delta z_{\psi\varphi} \chi_\alpha \varphi 
\end{equation} 
The $z$-factors in (\ref{fieldred}) and (\ref{fieldredgen})
are power series in the local coupling
\begin{equation}
\delta z_a \equiv \delta z_a(g^2) = \sum _{l\geq 1} \hat z_a^ {(l)}
g^{2l}(x) \ .
\end{equation}
 The corresponding counterterms
 are $\brs_{\Gacl}$-variations and are best expressed in
the form of symmetric differential 
operators acting on the classical action. These
symmetric operators are generalized field counting operators and
 can be found in their explicit form in  appendix B (\ref{Nanfang}) --
 (\ref{Nende}).

In general
the symmetric counterterms of loop order $l$ are a linear combination
of the two gauge independent counterterms (\ref{ctYM}) and (\ref{ctmatter})
and the counterterms of
field redefinitions (\ref{fieldred}) and (\ref{fieldredgen}). Here we
want to discuss the limit to constant coupling.
 For constant coupling it is possible to rewrite the matter
counterterm
(\ref{ctmatter}) into a mass and $q$-field
renormalization and
 field redefinitions of matter
fields, similarly the counterterm to the
Yang--Mills part can be rewritten into the  coupling renormalization and field
redefinitions and we obtain for constant coupling the following
invariant counterterms:
\begin{eqnarray}
\label{ctidentification}
\lim_{G \to g  } \Ga^ {(l)}_{\rm ct, inv} & = & 
\delta z_g^ {(1)} 
  g^ 2 \partial_{g^2}  + \delta z_A^ {(l)}( N_A + N_\rho
 + N_B +
N_{\cbar}
- 2 \xi \partial_\xi \bigr) \\
& &  { } + \delta z_\lambda^ {(l)} ( N_\lambda - N_{Y_\lambda}) +
+ \delta z_c^ {(l)} ( N_c - N_{Y_c}) \nonumber \\
& & { }+ \delta z_m^ {(l)} (N_q + N_{q ^\alpha} + N_{q_F} + m\partial_m )
\nonumber \\
& &{  } +  \delta z^{(l)} _\varphi (N_\varphi - N_{Y_\varphi})
+ \delta z^ {(l)}_\psi (N_\psi - N_{Y_\psi}) 
\Bigr)\Ga^ {\rm SYM}_{\rm
cl}\ . \nonumber 
\end{eqnarray}
In this expression the
non-renormalization theorems are reflected in the common gauge
independent $z$-factor for mass and $q$-field renormalization and in
the absence of higher-order renormalizations to the coupling constant.

For a unique definition of 1PI Green functions the free parameters
appearing
in (\ref{ctidentification}) 
have to be fixed by the  normalization
conditions on the residua of fields, on the matter mass term
 and by a normalization condition
on the coupling in one-loop order.  Normalization
conditions on the coupling $g$
in $l\geq 2 $ determine
 the coefficients  ${\cal O}(g^ 4)$ of the function
$\delta F(g^ 2) = r^{(1)}_\eta g^ 2 + {\cal O}(g^ 4)$ 
in the Slavnov--Taylor identity (\ref{STanomalous}). With a specific
choice for $\delta F(g^2 )$ the normalization of the coupling is
determined by the Slavnov--Taylor identity.

\section{The closed form of the gauge $\beta$-function
from the  anomalous breaking of supersymmetry}

The simplest application of the present construction 
 is the determination of the gauge
$\beta$-function. Since an invariant scheme does not exist the
$\beta$-functions cannot be inferred from the symmetric counterterms
or a local effective action,  but one has to construct the corresponding
partial differential equations, the Callan--Symanzik equation and the
renormalization group equation by algebraic consistency with the
anomalous Slavnov--Taylor operator (\ref{STanomalous}).

The Callan--Symanzik equation is the partial differential equation
connected with the breaking of dilatation. The  operator of dilatations
acts on the Green functions in the same way as a scaling of all mass
parameters of the theory according to their mass dimension:
\begin{equation}
{\cal W}^ D \Ga = 
- ( m\partial_m +\kappa \partial_\kappa) \Ga 
\equiv 
-\mu\partial_{\mu}  \ .
\end{equation}
$\kappa$ is a normalization point  introduced for defining infrared
divergent Green function off-shell. The variation of the scale
parameter $\kappa$ implies the renormalization group equation.

 The Callan--Symanzik equation continues the classical equation
\begin{equation}
\label{CSclassical}
\mu D_\mu \Gacl \equiv \mu \partial_{\mu} \Gacl - 
 m \intd \bigl(\frac {\delta} {\delta q} + \frac {\delta} {\delta
\qbar}\Bigr) \Gacl = 0
\end{equation}
to higher orders expressing the hard breakings of dilatations in form
of invariant operators. Hence, the Callan--Symanzik operator 
${\cal C} $ 
\begin{equation}
{ \cal C} =  \mu D_\mu + {\cal O}(\hbar)
\end{equation}
is a linear combination of the symmetric operators of the
theory and satisfies the consistency equations:
\begin{equation}
\label{CSSTcons}
\bigl( s^ {r_\eta}_{\Ga} + r ^{(1)} \delta {\cal S}\bigr) {\cal C }   \Ga = 
{\cal C}\bigl({\cal S}^ {r_\eta}(\Ga)+ r^ {(1)}\delta {\cal S} \Ga \bigr) +
\bigl(s^{r_\eta}_{\Ga} + r ^{(1)} \delta {\cal S}\bigr) \Delta_Y \ ,
\end{equation}
and 
\begin{equation}
\label{CSetacons}
\Bigl[ \Bigl(\frac {\delta } {\delta \eta}  - \frac {\delta} {\delta \etabar}
\Bigr), {\cal C} \Bigr] =  0 \ .
\end{equation}
In (\ref{CSSTcons}) the expression $\Delta_Y $ is  defined to be a collection of field
 monomials which are 
 linear in propagating fields. As such, $\Delta_Y$ appears  as a
trivial insertion  and does not need to  be absorbed into an
operator. The loop expansion of the Callan--Symanzik operator is a
 power series in the local coupling satisfying the topological formula
 (\ref{topfor}).

For the construction we first have to find the symmetric operators of
the classical Slavnov--Taylor operator ${\cal S}(\Ga)$
and have to extend them to
symmetric operators with respect to the anomalous Slavnov--Taylor operator
${\cal S} ^ {r_\eta}(\Ga) + r^ {(1)} {\delta S} \Ga$ (\ref{STanomalous}).

 When applied to the classical action the $\brs_\Ga$-invariant operators
 just have to reproduce the symmetric counterterms
constructed in the last section. Accordingly there are only two gauge
independent operators, one which corresponds to the one-loop
counterterm $\Ga_{\rm ct, YM}$ (\ref{ctYM}) and one corresponding to the $l$-loop
counterterms $\Ga_{\rm ct, matter}$ (\ref{ctmatter}).

$\Ga_{\rm ct, YM}$ is related to 
the one-loop renormalization of 
the local coupling and $\Ga_{\rm ct, matter}$ is related to a field
 redefinition of the axial vector multiplet into components of the
local supercoupling and the redefinition of 
 the $q$-multiplet (for details see \cite{KRST01})
\begin{eqnarray}
\label{YMoperator}
\Ga_{\rm ct, YM} & = &  g^ 4 \df {g^ 2} \Gacl + \mbox{field \ redef.}
 \\
\label{matteroperator}
\Ga_{\rm ct, matter} & = & {\cal D}_{Vv} \Gacl +  \mbox{field \ redef.}
\end{eqnarray}
The $\brs _\Ga$-invariant operator ${\cal D}_{Vv} $ is given by
\begin{eqnarray}
  {\cal D}^{(l)}_{Vv} 
& \equiv& \intd
 \biggl( v^ {{(G^{2l})} \mu} \frac {\delta} {\delta V^ \mu} +
\lambda^{{(G^{2l})}\alpha}
\frac {\delta} {\delta \lambdaV^\alpha}
+
\lambdaVbar ^ {{(G^ {2l})}\alphadot}
\frac {\delta} {\delta \lambdaVbar^\alphadot}
 \nonumber \\
& & { } \phantom{\intd}
+ d^ {(G^{2l})}  
\frac {\delta}{\delta \DV}
- i (\epsilon ^ \alpha \chi^ {(G^{2l})}_\alpha - \chibar^ {(G^{2l}) }
_\alphadot \epsilonbar^ \alphadot ) \frac {\delta}{\delta \tilde c} 
\nonumber \\
& & { }\phantom{\intd}
- 2 g^ {2l} (q +m) \frac {\delta }{\delta q} - 2 \bigl(g^{2l}
q^\alpha + 2 \chi^{{(G^{2l})}\alpha}(q+m)\bigr) 
\frac {\delta} {\delta q^ \alpha}
 \nonumber \\
& & { } \phantom{\intd}  -2
\bigl(g^ {2l}q_F + 2 f^ {(G^{2l})}(q+m)-  \chi^{{(G^{2l})}\alpha} q_\alpha\bigr) 
 \frac {\delta } {\delta q_F}
 \nonumber \\
& & { } \phantom{\intd}
- 2 g^ {2l} (\qbar +m) \frac {\delta }{\delta \qbar} - 2 \bigl(g^{2l}
\qbar^\alphadot +2 \chibar^{{(G^{2l})}\alphadot}(\qbar+m)\bigr) 
\frac {\delta} {\delta \qbar^ \alphadot}
 \nonumber \\
& & { } \phantom{\intd} -2
\bigl(g^{2l}\qbar_F
 + 2\fbar^ {(G^{2l})}(\qbar+m)-  \chibar^{(G^{2l})}_\alphadot \qbar
 ^\alphadot\bigr) 
 \frac {\delta } {\delta \qbar_F} \biggr)
\label{DVv}
\end{eqnarray}
In this expression the components of the  multiplet $G^
{2l}$ are defined by the following expansion of the local
supercoupling in superspace :
\begin{eqnarray}
\label{E2ldef}
G^{2 l}(x, \theta, \thetabar )  & = & \bigl({\mbox{\boldmath{$\eta$}}}(x,
\theta, \thetabar ) +{\mbox{\boldmath{$\etabar$}}}(x, \theta,
\thetabar )\bigr)^ 
{-l} \nonumber  \\
&\equiv &  g^{2l} (x) + \theta^ \alpha \chi^{(G^ {2l})}_\alpha +
 \chibar^{(G^{2 l})}_\alphadot \thetabar^ \alphadot + 
\theta^ 2 f^ {(G^ {2l})} +\thetabar^ 2 \fbar^ {(G^ {2l})} 
+ \theta \sigma^ \mu \thetabar v_\mu^ {(G^ {2l})} 
\nonumber  \\
& & { }
+i \theta^ 2 (\lambdabar^{(G^ {2l})}+
\frac 12 \partial_\mu \chi  \sigma^\mu)
 \thetabar 
-i \thetabar^ 2 \theta (\lambda^{(G^ {2l})} +
\frac 12  \sigma^\mu \partial_\mu \chibar^{(G^{2l})} )
\nonumber  \\
& & { }
   + \frac 14
\theta^ 2 \thetabar^2  (d^ {(G^ {2l}) } - \Box g^{2l}) 
\end{eqnarray} 

The two $\brs_\Ga$-invariant operators (\ref{YMoperator}) and
 (\ref{matteroperator})  have to be
 extended
 to symmetric operators with
respect to the anomalous Slavnov--Taylor operator ${\brs }^ {r_\eta}_\Ga
+ r^{(1)} {\delta S} $. 

For the operator $\intd g^ 4 \delta_{g^ 2}$ (\ref{YMoperator}) the
extension can be 
determined straightforwardly. It turns out that the operator
\begin{equation}
\label{Dkin}
{\cal D}_{\rm kin} =
\intd g^4 F(g^ 2) \df {g^2} = \intd g^ 4 (1 + r_\eta^ {(1)}g^2+ {\cal
O}(\hbar^ 2)) \df {g^2 }
\end{equation}
is symmetric with respect to the anomalous Slavnov--Taylor operator 
(\ref{STanomalous}).
And the function $F(g^2 )$ is the function modifying the classical
supersymmetry transformations of the coupling due to the anomaly
(\ref{brsanomalous}).

It is more involved to extend ${\cal D}_{Vv}$ 
(\ref{matteroperator}) to a symmetric operator
with respect to the anomalous Slavnov--Taylor operator.
The corresponding operator for $r_\eta = 0$ has been already
constructed in SQED \cite{KRST01}.
 From this operator the extension  can be determined
straightforwardly:  In a first step we extend ${\cal D}_{Vv}$ to a 
$\brs ^{r_\eta} _{\Ga} $-invariant operator by replacing the components
of the local supercoupling in (\ref{DVv})
by the components of the local
supercoupling with anomalous supersymmetry transformations:
\begin{equation}
\label{DvVreta}
 {\cal D}_{Vv}(G^ {2l}) \to {\cal D}^{r_\eta} _{Vv}\equiv  {\cal
 D}_{Vv}(\tilde G^ {2l})
\end{equation}
and the components of $\tilde G$ are determined by the anomalous supersymmetry
transformations (\ref{brsanomalous})
\begin{eqnarray}
\label{tildeG}
\delta^ {r_\eta}_\alpha \tilde G^{2l} = \Bigl( \frac
{\partial}{\partial \theta^ \alpha} 
+ i (\sigma \thetabar)\alpha \partial_\mu \Bigr)\tilde G^ {2l} \ ,\\
\bar \delta^ {r_\eta}_\alphadot \tilde G^{2l} =
- \Bigl( \frac {\partial}{\partial \thetabar^
\alphadot}
+ i (\theta \sigma )_\alphadot \partial_\mu \Bigr) \tilde G^ {2l} \ .\nonumber
\end{eqnarray}

For the further calculation only 
the lowest components of the anomalous
multiplet of the supercoupling are relevant:
\begin{eqnarray}
\chi^{(\tilde G^ {2l})}_\alpha & = & - l g^ {2(l+1)} F(g^2) \chi_\alpha \ ,
\qquad
\chibar^{(\tilde G^ {2l})}_\alphadot = - l g^ {2(l+1)}  F(g^2)
\chibar_\alphadot\ , 
 \\
v_\mu^ {(\tilde G^ {2l})} &  = & l g^ {2(l+1)}\Bigl( F(g^2)
i \partial_\mu (\eta - \etabar)  
+ \frac{ g^ 2}2  \bigl((l+1)F(g^2) + g^ 2 F'(g^2) \bigr)
   \chi \sigma_\mu \chibar \Bigr). \nonumber 
\label{vtildeE2l}
\end{eqnarray}

The $\brs_\Ga^ {r_\eta}$-invariant operator ${\cal D}_{Vv}^{r_\eta}$
(\ref{DvVreta})
 can be extended to a symmetric operator ${\cal D}^ {\rm
sym}_{Vv}$ in the same way as in SQED. For proving consistency of the
method we have determined the complete symmetric operator. For 
 constant coupling 
 only the following commutator is relevant:
\begin{eqnarray}
& &\bigl[{r^{(1)} \delta {\cal S}}, -i \intd (\epsilon\chi^ {(\tilde E^{2l})} -
\chibar^ {(\tilde E^{2l})}\epsilonbar)\df {\tilde c} ) \bigl] \Big|
_{V= 0\atop \tilde c = 0} \\ 
& & =
\intd (\epsilon\chi^ {(\tilde E^{2l})} -
\chibar^ {(\tilde E^{2l})}\epsilonbar)(\df \eta - \df \etabar ) \ .
\end{eqnarray}
It can be compensated by a differential operator with respect to
the coupling:
\begin{equation}
\label{Dg}
D^ {(l+1)}_{g^2} = \intd \bigl( 
F(g^2) g^ {2(l+2)} \df {g^2} +  \chi^ {(\tilde G^{2l})} \df \chi + 
\chibar^ {(\tilde G^{2l})} \df \chibar + \cdots \bigr) \ .
\end{equation}
 and results in  a contribution to the $\beta$-function of the coupling
generated by  the Adler--Bardeen anomaly.

The complete operator $   {\cal D}^ {\rm sym}_{Vv}$ 
\begin{equation}
\label{DvVsym}
  {\cal D}^ {\rm sym}_{Vv} \equiv
 {\cal D}^ {(l)}_{Vv} - 8 r^ {(1)}\bigl(   {\cal D}^ {(l+1)}_{g^2} +
 l (  {\cal N}^ {(l+1)}_V - 8 (l+1)r ^{(1)} \delta  {\cal N}^ {(l+2)}_V )
 \bigr) ,
\end{equation}
contains in addition a field operator for the axial vector multiplet
${\cal N}_V$ 
which also depends on the function $F (g^2 )$. The complete operator
is given in the appendix B.

All further symmetric operators of the Callan--Symanzik operator are
gauge dependent. 
 They correspond to the field redefinitions (\ref{fieldred}) and
(\ref{fieldredgen}) and can be
constructed  as $\brs^ {r_{\eta}}_\Ga + r^{(1)} \delta {\cal
S}$-variations of local field monomials (\ref{Nanfang}) -- (\ref{Nende}).
In the limit to constant coupling they are the usual field counting
operators.

Altogether  the Callan--Symanzik operator is a linear combination of
the two gauge independent operator ${\cal D}_{\rm kin}$
(\ref{Dkin}) and ${\cal D}^ {\rm sym}_{Vv}$(\ref{DvVsym})
and the gauge dependent operator ${\cal N}_a$ (\ref{Nanfang}) -- (\ref{Nende}):
\begin{eqnarray}
\label{CSoperator}
{\cal C}& = & \mu_i D _{\mu_i} +
\hat \beta_{g^2}^ {(1)} {\cal D}_{\rm kin}+ 
\sum_{l} \Bigl(  \gamma^ {(l)} {\cal D}^ {{\rm sym}{(l+1)}}_{Vv}
- \sum_a \hat\gamma_a^ {(l)} {\cal N}^{(l)}_a \Bigr) \ .
\end{eqnarray}
Here the sum over $a$ includes all possible field redefinitions.
The algebraic construction yields the Callan--Symanzik equation 
supersymmetric non-abelian gauge theories with local gauge coupling
and  with the axial vector multiplet:
\begin{equation}
\label{CSequation}
{\cal C} \Ga = \Delta_Y \ .
\end{equation}
It is valid for all 1PI Green functions involving propagating fields
and describes also the scale transformations of axial current Green
functions in presence of the anomaly.

In its structure the Callan--Symanzik equation of non-abelian gauge
theories is the same as the one of SQED, in particular there is only a
one-loop independent coefficient for the renormalization of the gauge
coupling.
  The difference concerns 
the explicit expressions depending here on the non-vanishing anomaly
coefficient $r^{(1)}_\eta$ via the function $F(g^2)$.

Here we are mainly interested in the limit to constant coupling:
For constant coupling 
we get  from (\ref{CSoperator}) and (\ref{CSequation}) the Callan--Symanzik equation
of  supersymmetric non-abelian gauge theories:
\begin{eqnarray}
\label{CSconst}
& & \Bigl(\mu_i\partial_{\mu _i} +
g^4 F(g^2) (\hat \beta^ {(1)}_{g^2} + 8 r^ {(1)} \gamma) \partial_{g^2}
- \sum_a \gamma_a {\hat {\cal N}}_a )\Ga^ {\rm SYM} \nonumber \\
& & = (1-2\gamma) \intd m \bigl(\frac{\delta}{\delta q} +
\frac{\delta}{\delta \qbar}
\bigr)\, \Ga\Big|_{q^i = 0\atop V^ i = 0} + \Delta_Y \ .
\end{eqnarray}
and  the $\beta$-function is
determined by the operator $\beta_{g^2}^ {(1)}{\cal D}_{\rm kin} +
8 r^ {(1)} \hat \gamma {\cal D}_{g^2} $
 in a closed form:
\begin{equation}
\beta_{g^2} = (\hat \beta_{g^2}^ {(1)} + 8 r^ {(1)} \gamma ) g^ 4 F(g^2) 
\quad \mbox{with} \quad 
\gamma = \sum_l g^ {2l} \hat \gamma ^ {(l)} \ .
\end{equation}
Thus, in $l\geq 2$ the gauge $\beta$-function of order $l+1$ 
is completely determined
from a complete $l$-loop calculation. This amounts  to determine the
order $l$  of $F(g^2)$ in the Slavnov--Taylor identity and the $l$-loop 
anomalous mass dimension.
In particular, the two-loop order is determined by the anomaly
coefficient $r^{(1)}_\eta$, the coefficient of the Adler--Bardeen anomaly
$r^ {(1)}$ and the scheme independent coefficient $\gamma^ {(1)}$.
Higher orders are scheme dependent.

 When we choose in the Slavnov--Taylor identity the function $F(g^2) =
\frac 1 {1- r^{(1)}_\eta g^2}$ (\ref{FNSZV})
 we get  the
NSVZ-formula of the gauge $\beta$-function \cite{NSVZ83}:
\begin{equation}
\label{NSVZ}
\beta^ {\rm NSVZ}_{g^2} = (\hat \beta_{g^2}^ {(1)} + 8  r^ {(1)} \gamma ) g^ 4 
\frac 1 {1- r^{(1)}_\eta g^2}\ .
\end{equation}
For the choice  $F(g^2) = 1 + r^{(1)}_\eta g^2 $ (\ref{F2loop})
we obtain 
a purely two-loop function in  Super--Yang--Mills theories without matter:
\begin{equation}
\label{beta2loop}
\beta^{\rm min}_{g^2} = (\hat \beta_{g^2}^ {(1)} + 8 r^ {(1)} \gamma ) g^ 4 
  {(1+ r^{(1)}_\eta g^2)}\ .
\end{equation}

The renormalization group equation can be determined in the same way
as the Callan--Symanzik equation and consists of the same number and
same types of
symmetric
operators implying the same relation among them in the limit to
constant coupling. The $\beta$-functions of the renormalization group
equation
 are scheme dependent already in one-loop
order. To find the relations to the $\beta$-functions of the
Callan--Symanzik equation, which are
usually quoted, one has to take
asymptotic normalization conditions, which result in a trivial
equation for the scaling of the supersymmetric mass parameter \cite{KR94,KRST01soft}:
\begin{equation}
m\partial_ m \Ga  
 =  m \intd  \bigl(\frac{\delta}{\delta q} +
\frac{\delta}{\delta \qbar}
\bigr)\, \Ga\ .
\end{equation}
With this equation
it is possible  to eliminate the field
differentiation $\delta _q $ and $\delta_\qbar$ in (\ref{CSconst}) for
constant coupling and one obtains
 the renormalization group equation for asymptotic normalization
conditions:
\begin{equation}
\bigl(\kappa_\infty \partial_{\kappa_\infty} + 
g^4 F(g^2) (\hat \beta^ {(1)}_{g^2} + 8 r^ {(1)} \gamma) \partial_{g^2}
+  2 \gamma m \partial_m 
- \sum_a \gamma_a {\hat {\cal N}}_a \bigr)\Ga^ {\rm SYM} = 0 \ ,
\end{equation}
It is a homogeneous differential equation for the 1PI vertex
functional of susy gauge theories with constant coupling, and it
contains the same
coefficient functions as the Callan--Symanzik equation.
The renormalization group equation can be integrated and defines a running gauge coupling
to higher orders. It is tempting to use an all order expression
as (\ref{NSVZ}) or (\ref{beta2loop}) for the integration, 
however, an interpretation of the corresponding running coupling
is  meaningless, as long as it is not related to specific
normalization properties of the 
 coupling.

\section{Conclusions}

 In the present paper
we  have continued the algebraic analysis of non-renorm\-al\-ization theorems
via local couplings  \cite{FLKR00,KRST01}
to supersymmetric Yang--Mills theories. As a result of the construction one
obtains 
the generalized non-renormalization theorem of the gauge coupling,
which gives rise to a pure one-loop $\beta$-function. But in addition
we find
an anomalous breaking of supersymmetry in one-loop order, and 
 it is the supersymmetry anomaly which is
responsible for the two-loop term of the gauge $\beta$-function
in supersymmetric pure Yang-Mills
theories.

 Higher-order terms of the gauge $\beta$-function
   depend on the  normalization conditions for the coupling.
   But if one performs a complete $l$-loop
calculation in Super--Yang--Mills with local coupling,
 the gauge $\beta$-function of order
$l+1$ 
is unambiguously
determined in terms of lower order coefficients. Such coefficients are 
in particular
the  higher-order modifications of  supersymmetry transformations in the
Slavnov--Taylor identity and we find
 the conditions for the NSVZ expression of the $\beta$-function as
conditions on the supersymmetry transformation of the gauge coupling.

Hence, one can now derive
the closed expression of the gauge $\beta$-function  which has been
first found  in the framework of
the instanton calculus \cite{NSVZ83} in a purely perturbative and even
scheme independent framework.  

Without local coupling the improved renormalization properties alias the
non-renormalization theorems notoriously used to escape a rigorous
treatment in  perturbation theory, and just the same happened
 for the closed expression of the gauge
$\beta$-function.  One
way to find restrictions on the $\beta$-function is the usage of the
supercurrent with its anomalies. From this construction one finds a
similar but not as restrictive expression of the gauge
$\beta$-function \cite{CPS79,LPS87} as in the instanton calculus, which
takes into account possible redefinitions suppressed in the latter
derivation but found in the derivation with local couplings again. The
construction of the supercurrent is beyond the scope of the present
paper, but the supercurrent and its anomaly multiplet
is  an important topic, which should be
reanalyzed with the local coupling again.

Another perturbative  approach to the closed expression of the
  gauge $\beta$-function  is based on
  holomorphicity
 of the Wilsonian effective action \cite{SHVA86,SHVA91, SEI93}.  
Introducing the notion of a
Wilsonian coupling
 the Wilsonian 
effective action of the Yang-Mills part is exhausted in one-loop
order and one obtains indeed 
 an indication of the generalized non-renormalization theorem
for the gauge coupling. 
To explain the 
 higher-order terms in the gauge $\beta$-function one has used 
again the construction of 
   the supercurrent and the Adler--Bardeen anomaly in its supersymmetric
  extension \cite{CPS79,KO81},
 but
the higher-order terms of  the $\beta$-function in pure Super--Yang--Mills
remain mysterious in the Wilsonian approach.
There are attempts to explain them by  passing over from  
the Wilsonian coupling in the Wilsonian
effective action to a canonical coupling
constant by rescaling the vector fields \cite{ARMU97}, but
 in effect such redefinitions would be
  scheme dependent.

Holomorphicity as used in the Wilsonian approach
 can be used in its full extent only if the coupling is 
treated as a space-time dependent external superfield
 and even then, holomorphicity is not a principle of
quantum field theory, but can be valid at most for some local parts of
 the full Green functions.
The Wilsonian coupling can be compared to the chiral and the
antichiral fields
composing the gauge supercoupling in the present paper, however,
 these fields are here introduced in a completely scheme independent framework.
 Performing renormalization in presence of the space--time  dependent
gauge supercoupling  consistently to its very end, holomorphicity of
 symmetric counterterms is a consequence of classical symmetries and
 non-holomorphicity of the gauge $\beta$-function arises from the two
 anomalies of these classical symmetries, from
 the
 anomaly of supersymmetry found in the present paper and from  the 
Adler--Bardeen anomaly of the axial current. Hence, the
 non-holomorphic dependence of
 the
gauge 
$\beta$-function is deeply related to the failure of invariant
schemes and of a naive application of multiplicative renormalization
via $z$-factors in presence of the local gauge coupling.

\vspace{0.5cm}
{\bf Acknowledgments}

We thank R. Flume  and D. St\"ockinger  
for many valuable discussions.
We are grateful to the Max--Planck--Institut f\"ur Physik, Munich, 
 for kind hospitality where  the final version of the paper has been
  worked out.

\newpage

\begin{appendix}
\section{The BRS transformations} 
 
In this appendix we list the BRS transformations of all fields
introduced in classical action with local gauge coupling and with 
the axial vector multiplet. 
\begin{itemize}
\item BRS transformations of the gauge vector multiplet
\begin{eqnarray}
\brs A_\mu & = & \frac 1g \partial_\mu g c + i \big[ A_\mu, c\big]
 + i\epsilon\sigma_\mu\lambdabar
             -i \lambda\sigma_\mu\epsilonbar \\
 &  &  +
\frac 12 g ^2(\epsilon \chi + \chibar
\epsilonbar)A_\mu-i\omega^\nu\partial_\nu A_\mu 
\ ,\nonumber \\
\brs\lambda^\alpha & = & - i \bigl\{ \lambda, c\bigr\} +
\frac{i}{2g} (\epsilon\sigma^{\rho\sigma})^\alpha
             G_{\rho\sigma}(gA) + i\epsilon^\alpha\,
             g(\phibar_L T_a \varphi_L - \varphi_R T_a \phibar_R) \tau_a
 \nonumber \\
& & + \frac 12 \epsilon^ \alpha
g^2 ( \chi \lambda - \chibar \lambdabar) 
+ \frac 12 g ^2(\epsilon \chi + \chibar \epsilonbar) \lambda^ \alpha
              -i\omega^\nu\partial_\nu  
             \lambda^\alpha
\ ,\nonumber \\
\brs \lambdabar_\alphadot & = & - i \bigl\{ \lambdabar, c\bigr\} +
\frac{-i}{2g} (\epsilonbar\sigmabar^{\rho\sigma})
             _\alphadot G_{\rho\sigma}(gA) + i g \epsilonbar_\alphadot\, 
             (\phibar_L T_a \varphi_L-\varphi_R T_a \varphi_L) \tau_a
\nonumber \\
& &  + \frac 12 \epsilonbar_ \alphadot
g^2 ( \chi \lambda - \chibar \lambdabar)
+ \frac 12 g ^2(\epsilon \chi + \chibar \epsilonbar) \lambdabar_ \alphadot
             -i\omega^\nu\partial_\nu \lambdabar_\alphadot \ . \nonumber
\end{eqnarray}
\item BRS transformations of the axial vector multiplet
\begin{eqnarray}
\brs V_\mu & = & \partial_\mu \tilde c + i\epsilon\sigma_\mu{\bar {\tilde
\lambda}}
             -i \tilde
\lambda\sigma_\mu\epsilonbar -i\omega^\nu\partial_\nu V_\mu
\ ,\\
\brs\tilde\lambda^\alpha & = & \frac{i}{2} (\epsilon\sigma^{\rho\sigma})^\alpha
             F_{\rho\sigma}( V) + \frac i2 \epsilon^\alpha\,
             \tilde D -i\omega^\nu\partial_\nu  
             \tilde \lambda^\alpha
\ ,\nonumber \\
\brs{\bar {\tilde \lambda}}
_\alphadot & = & \frac{-i}{2} (\epsilonbar\sigmabar^{\rho\sigma})
             _\alphadot F_{\rho\sigma}(V) + \frac i 2 \epsilonbar_\alphadot 
\tilde D
             -i\omega^\nu\partial_\nu \bar {\tilde \lambda}_\alphadot
\ , \nonumber \\
\brs \tilde D & = & 2 \epsilon \sigma^ \mu \partial_\mu {\overline {\tilde \lambda}} + 
                  2 \partial_
 \mu  {\tilde \lambda} \sigma^ \mu \epsilonbar 
             -i\omega^\nu\partial_\nu  {\tilde D} \ .\nonumber
\end{eqnarray}
\item BRS transformations of  matter fields 
\begin{eqnarray}
\brs\varphi_L & = & i(g c_a T_a - \tilde c)\varphi_L +\sqrt2\,
\epsilon\psi_L
 - i\omega^\nu\partial_\nu \varphi_L
\ , \\
\brs\phibar_L & = & -i \phibar_L (g c_a T_a - \tilde c)\,   +
\sqrt2\, \psibar_L\epsilonbar - i\omega^\nu\partial_\nu \phibar_L
\ ,\nonumber \\
\brs\psi_L^\alpha & = & i(g  c_a T_a - \tilde c)\,\psi_L^\alpha
\nonumber \\
 & &  - \sqrt2\, 
         \epsilon^\alpha\, (\qbar+m)\phibar_R
         -\sqrt2\, i (\epsilonbar\sigmabar^\mu)^\alpha D_\mu\varphi_L 
         -i\omega^\nu\partial_\nu \psi_L^\alpha
\ ,\nonumber \\
\brs{\psibar_L}_\alphadot & = & i{\psibar_L}_\alphadot (g T_r c_r - \tilde c)\,
\nonumber \\
& &          + \sqrt2\,\epsilonbar_\alphadot\,(q+ m)\varphi_R + \sqrt2\, i
         (\epsilon\sigma^\mu)_\alphadot D_\mu\phibar_L
         -i\omega^\nu\partial_\nu {\psibar_L}_\alphadot \ .\nonumber
\end{eqnarray}
and respective expressions for right-handed fields.
\item The BRS transformations of ghosts
\begin{eqnarray}
\brs c & = & - ig c \, c  + 2i\epsilon\sigma^\nu\epsilonbar A_\nu +
\frac 12 g^2 (\epsilon \chi + \chibar \epsilonbar) c
-i\omega^\nu\partial_\nu c
\ ,\\
\brs\tilde  c & = & 2i\epsilon\sigma^\nu\epsilonbar V_\nu
-i\omega^\nu\partial_\nu\tilde  c
\ ,\nonumber \\
\brs\epsilon^\alpha & = & 0
\ ,\nonumber \\
\brs\epsilonbar^\alphadot & = &0
\ ,\nonumber \\
\brs\omega^\nu & = & 2\epsilon\sigma^\nu\epsilonbar \ . \nonumber
\end{eqnarray}
\item BRS transformations of the $B$-field, anti-ghost and gauge parameter
\begin{eqnarray}
\label{BRSgaugefixing}
\brs B &  = &
2i \epsilon \sigma ^ \nu \epsilonbar \partial_\nu 
\cbar - i
\omega^ \nu
\partial_ \nu B \ ,\\
\brs \cbar &  = & B
  - i
\omega^ \nu
\partial_ \nu \cbar \ , \nonumber \\
\brs \chi_\xi &  = &
2i \epsilon \sigma ^ \nu \epsilonbar \partial_\nu
\xi   - i
\omega^ \nu
\partial_ \nu \chi_\xi \ ,\nonumber \\
\brs \xi &  = & \chi_\xi
 - i
\omega^ \nu
\partial_ \nu \xi \ ,\nonumber \\
\brs C^ \mu &  = &
2i \epsilon \sigma ^ \nu \epsilonbar \partial_\nu
H^ \mu   - i
\omega^ \nu
\partial_ \nu C^ \mu \ ,\nonumber \\
\brs H^ \mu &  = & C^ \mu
 - i
\omega^ \nu
\partial_ \nu H^\mu \ . \nonumber
\end{eqnarray}
\item BRS transformations of the local coupling and its superpartners (\ref{E2def})
\begin{eqnarray}
\label{BRScoupling}
\brs g^2 & = &-( \epsilon^ \alpha \chi_\alpha + \chibar_\alphadot
\epsilonbar^ \alphadot) g^4   -i \omega ^ \nu \partial_\nu g^2 \ ,\\
\brs (\eta -\etabar) & = & (\epsilon^ \alpha \chi_\alpha
-\chibar_\alphadot \epsilonbar^ \alphadot )
 -i \omega ^ \nu \partial_\nu(\eta -
\etabar) \ ,\nonumber \\
\brs \chi_\alpha& = & -  i (\sigma^ \mu \epsilonbar)_\alpha ( \frac 1 {g^4}
\partial_\mu g^2  - \partial_\mu(\eta - \etabar)) + 2
\epsilon_\alpha f  
- i \omega^ \mu \partial_\mu
\chi_\alpha \ ,\nonumber \\
\brs \chibar_\alphadot& = & - i (\epsilon \sigma^ \mu)_\alphadot
( \frac 1 {g^4}
\partial_\mu g^2  + \partial_\mu(\eta - \etabar))  - 2
\epsilonbar_\alphadot \bar f 
- i \omega^ \mu \partial_\mu
\chibar_\alphadot \ ,\nonumber \\
\brs f & = &   i 
\partial_\mu \chi \sigma^ \mu \epsilonbar - i \omega^ \mu \partial_\mu
f \ ,\nonumber \\
\brs \bar f & = & - i 
 \epsilon \sigma^ \mu \partial_\mu \chibar 
- i \omega^ \mu \partial_\mu
\bar f \ . \nonumber
\end{eqnarray}
\item BRS transformations of $q$-multiplets (\ref{qdef})
\begin{eqnarray}
\brs q & = &  + 2i \tilde c (q  + m ) +
 \epsilon^ \alpha q_\alpha -i \omega ^ \nu \partial_\nu q\ ,\\
\brs \qbar & = & -2i \tilde c (\qbar + m)
+ \qbar_\alphadot \epsilonbar^ \alphadot
 -i \omega ^ \nu \partial_\nu
\qbar \ , \nonumber \\
\brs q_\alpha& = & + 2i \tilde c q_\alpha +2 i (\sigma^ \mu \epsilonbar)_\alpha 
D_\mu (q +m) + 2
\epsilon_\alpha q_F 
- i \omega^ \mu \partial_\mu
q_\alpha \ ,\nonumber \\
\brs \qbar_\alphadot& = &  - 2i \tilde c \qbar_\alphadot +
2 i (\epsilon \sigma^ \mu)_\alphadot
 D_\mu (\qbar +m )   - 2
\epsilonbar_\alphadot \qbar_F
- i \omega^ \mu \partial_\mu
\qbar_\alphadot \ ,\nonumber \\
\brs q_F & = &+ 2i \tilde c q_F +
 i 
D_\mu q^ \alpha \sigma^ \mu_{\alpha \alphadot} \epsilonbar^ \alphadot
- 4 i
 \lambdaVbar _ \alphadot \epsilonbar^ \alphadot (q +m )
 - i \omega^ \mu \partial_\mu
q_F \ , \nonumber \\
\brs \qbar_F & = & - 2i \tilde c \qbar_F  - i 
 \epsilon^ \alpha \sigma^ \mu_{\alpha \alphadot} D_\mu \qbar^ \alphadot 
+ 4 i
\epsilon ^ \alpha \lambdaV _ \alpha  (\qbar + m)
- i \omega^ \mu \partial_\mu
\qbar_ F \ . \nonumber
\end{eqnarray}
\end{itemize}

Supplementing the classical action by the external field part,
\begin{eqnarray}
\label{extf}
\Gamma_{\rm ext} & = &  \intd \Bigl(\rho_a^ \mu \brs A_{\mu a} +
Y_{\lambda a}^\alpha \brs\lambda_{\alpha a}
+ Y_\lambdabar{}_{\alphadot a} \brs\lambdabar^\alphadot_a + Y_{c a} \brs c_a
\nonumber\\&&{}\hfill
+ Y_{\varphi_L} \brs\varphi_L + Y_{\phibar_L} \brs\phibar_L
+ Y_{\psi_L}^\alpha \brs\psi_L{}_\alpha
+ Y_{\psibar_L}{}_\alphadot \brs\psibar_L^\alphadot
+ ({}_{L\to R}) \nonumber\\&&{ }\hfill
+  \frac 12(Y_{\lambda a}\epsilon-\epsilonbar
                     Y_{\lambdabar a})
(Y_{\lambda a}\epsilon-\epsilonbar
                     Y_{\lambdabar a})
                   -2\bigl((Y_{\psi_L} \epsilon)(\epsilonbar
Y_{\psibar_L}) +
({}_{L\to R})             \Bigr)
 \  ,
\end{eqnarray}
the classical Slavnov--Taylor identity (\ref{STclassical})
expresses in functional form BRS invariance of the classical action
and on-shell  nilpotency of BRS transformations.  
 The Slavnov--Taylor operator acting on a general
functional ${\cal F}$  is defined as
\begin{eqnarray}
{\cal S}({\cal F}) & = & 
\intd\biggl(\Tr \Bigl(\dF{\rho_{\mu}} \dF{A^\mu}
+ \dF{Y_{\lambda }{}_\alpha}\dF{\lambda^\alpha}
+ \dF{Y_{\lambdabar }^\alphadot}\dF{\lambdabar_{\alphadot }}
 \nonumber\\&&{}\quad
+   \dF{Y_{c }} \dF{c}         +\brs B \dF{B} +\brs\bar{c}
\dF{\bar{c}}\Bigr)
 \nonumber\\&&{}\quad
 +\brs
\xi \dF{\xi} 
+ \brs\chi_\xi \dF{\chi_\xi}  +\brs
H^\mu \dF{H^\mu} 
+ \brs C_\mu \dF{C^\mu} 
\nonumber\\&&{}\quad
+ \dF{Y_{\varphi_L}}\dF{\varphi_L}
+ \dF{Y_{\phibar_L}}\dF{\phibar_L}
+ \dF{Y_{\psi_L{}_\alpha}}\dF{\psi_L^\alpha}
+ \dF{Y_{\psibar_L}^\alphadot}\dF{\psibar_L{}_\alphadot}
+(_{L\to R})\Bigr)
\nonumber\\&&{} \quad
+\brs G^i\frac{\delta{\cal F}}{\delta G^ i}
+\brs V^i\frac{\delta{\cal F}}{\delta V^ i}
+\brs q^i\frac{\delta{\cal F}}{\delta q^ i}
+\brs\qbar^i\frac{\delta{\cal F}}{\delta\qbar^ i}
 \biggr)
+\brs\omega^\nu \frac{\partial{\cal F}}{\partial\omega^\nu} \ .
\label{STOperator}
\end{eqnarray}
Here $G^i $ summarizes the components of the local supercoupling,
\begin{equation}
G^ i = (g^2, \eta-\etabar, \chi, \chibar, f , \fbar)
\end{equation}
$V^i$ the fields of the axial vector multiplet and $q^i$ and 
$\qbar^ i$ the fields of
the chiral and antichiral multiplet.

The Slavnov--Taylor operator and its linearized version $\brs_\Ga$ have
the following nilpotency properties:
\begin{eqnarray}
\label{nilpotency}
& & \brs_{\cal F} {\cal S} ({\cal F}) = 0 \quad \mbox{for any functional}\
{\cal F} \ ,\\
& & \brs_{\cal F} \brs_{\cal F} = 0\ ,  \quad \mbox{if} \quad {\cal S}
({\cal F}) = 0\nonumber \ . 
\end{eqnarray}

\newpage

\section{Symmetric operators}

In the Wess--Zumino gauge the symmetric operators 
 of Super--Yang--Mills theories with local gauge coupling 
are  symmetric 
  with
respect to the anomalous Slavnov--Taylor identity
(\ref{CSSTcons}), the $\eta-\etabar$ identity (\ref{CSetacons}) and
 satisfy order by order the topological formula (\ref{topfor}). 

\begin{itemize}
\item The gauge independent differential operator 
of the gauge coupling ${\cal D}_{kin}$ (\ref{Dkin}), which gives rise
to the one-loop $\beta$ function:  
\begin{eqnarray}
\label{Dkinapp}
{\cal D}_{\rm kin}
 &= & \intd g^ 4 F(g^2)
 \frac{\delta} {\delta g^2}
\end{eqnarray}
\item The gauge independent operator ${\cal D}^ {\rm sym}_{Vv}$, which 
extends
 the  $\brs^ {r_\eta}_\Ga$ invariant operator ${\cal D}^{r_\eta}_{Vv}$
(\ref{DvVreta})
to a 
$\brs^{r_\eta}_\Ga + r ^{(1)}\delta {\cal S}$-symmetric operator:
\begin{equation}
\label{DvVsym2}
{\cal D}^ {\rm sym}_{Vv} \equiv
{\cal D}^ {r_\eta\,(l)}_{Vv} - r^ {(1)}\bigl( 8 {\cal D}^ {(l+1)}_{g^2} +
8 l ( {\cal N}^ {(l+1)}_V - 8 (l+1)r ^{(1)} \delta {\cal N}^ {(l+2)}_V )
 \bigr) .
\end{equation}
with
\begin{eqnarray}
\label{Dgapp}
{\cal D}^ {(l+1)}_{g^2} & = & 
 \intd \biggl( F(g^2) g^ {2(l+2)}\frac {\delta }{\delta g^2} -  
\chi^ {(\tilde G^ {2l})\alpha} \frac {\delta }{\delta \chi^ \alpha} +
\chibar^ {(\tilde G^ {2l})\alphadot} \frac {\delta }{\delta \chibar^
\alphadot} \nonumber \\
& & \phantom{ 4 \intd} - 
f^ {(\tilde G^ {2l})} \frac {\delta }{\delta f} -
\fbar^ {(\tilde G^ {2l})}
 \frac {\delta }{\delta \fbar} \biggr) \ ;
\end{eqnarray}
\begin{eqnarray}
\label{NV}
{\cal N}^ {(l)}_V & = &
 \intd \biggl( F(g^2) g^ {2l}\Bigl( V^ {\mu} \frac {\delta }{\delta V^ \mu} +
 \lambdaV^ {\alpha} \frac {\delta }{\delta \lambdaV^ \alpha} +
 \lambdaVbar^ {\alphadot} \frac {\delta }{\delta \lambdaVbar^ \alphadot} +
 \tilde D  \frac {\delta }{\delta \tilde D}\Bigr) \nonumber \\
& & \qquad - (F(g^2) + \frac {g^2}{l-1} \partial_{g^2} F(g^2))
\bigl(\frac i 2 V_\mu (\sigma^ \mu \chibar^ {(\tilde G ^{2l})})^ \alpha \frac {\delta }{\delta
 \lambdaV^ \alpha} + \mbox{c.c.}
  \nonumber \\
& &  \qquad \qquad - 2 \Bigr( V^ \mu v_{\mu}^
 {(\tilde G^ {2l})}  - i \lambdaV^ \alpha \chi^ {(\tilde G^ {2l})}_\alpha 
+ i\chibar^ {(\tilde G^ {2l})}_ \alphadot\lambdaVbar^ \alphadot 
 \Bigl)
\frac {\delta }{ \delta \tilde D }\bigr)  \biggr) \ ,
\end{eqnarray}
and 
\begin{eqnarray}
\delta {\cal N}^ {(l+1)}_V
 & = &  \intd g^ {2(l+1)} F(g^2)\bigl( F(g^2) + \frac {g^2} l
 \partial_{g^2} F(g^2)\bigr) 
 V^ \mu V_\mu \frac {\delta }{\delta \tilde D }\ .
\end{eqnarray}
In eqs.~(\ref{NV}) and (\ref{Dgapp})
   the components of the  anomalous multiplet $\tilde G^
{2l}$ of the gauge coupling are defined in (\ref{tildeG}).
\item The symmetric differential operators of the gauge vector multiplet
\begin{eqnarray}
{\cal N}^{(l)}_A \Ga + \Delta^{(l)}_{Y, A}& = &
\tilde\brs_\Ga  \Bigl(\Tr \intd g^
{2l} 
f_A(\xi)\rho^\mu A_\mu \label{Nanfang} \\
& & \phantom{\tilde\brs_\Ga} -  \Tr \intd g^
{2l} f_A(\xi) \bigl( \cbar \df B - 2 \xi
 \df \chi - 
 \partial _\mu \df {C^ \mu}\bigr)\Bigr) \Ga \nonumber\\
{\cal N}^{(l)}_\lambda \Ga + \Delta^{(l)}_{Y, \lambda} & = &
\tilde \brs_\Ga \Tr \intd g ^{2l}f_\lambda(\xi)
(\lambda^ \alpha Y_{\lambda \alpha} +
\lambdabar_\alphadot Y^ \alphadot_{\lambdabar}) \nonumber \\
{\cal N}^{(l)}_{\lambda A} \Ga + \Delta^{(l)}_{Y, \lambda A} & = &
\tilde \brs_\Ga \Tr \intd i g^ {2(l+1)}f_{\lambda A}(\xi)
( Y_{\lambda}^ { \alpha} \sigma^
\mu_{\alpha
\alphadot} \chibar^ \alphadot +
\chi^ \alpha \sigma_{\alpha \alphadot}  Y^ \alphadot_{\lambdabar})
A_\mu \nonumber 
\end{eqnarray}
with
\begin{equation}
\tilde \brs_\Ga \equiv 
\brs^{r_\eta} _\Ga + r ^{(1)}\delta {\cal S} 
\end{equation}
In the expression for the field redefinition of the vector field
$A^ \mu$ we have used that the gauge fixing part is not renormalized due
to linearity of the $B$-field. Therefore the renormalizations of the
$B$-fields, the ghosts $\bar c$ and the gauge parameter doublet are
determined by the renormalization constants of the vectors. Working
out the variations we find the following invariant operators:
\begin{eqnarray}
{\cal N}^{(l)}_A & = &  \intd g^{2l} f_A(\xi)
\Bigl(A_a^\mu\df{A_a^\mu} - \rho_a^\mu\df{\rho_a^\mu}
 - B_a
\df {B _a}
- \cbar_a\df {\cbar_a} \\
& & \phantom{\intd g^{2l}f_A(\xi) }
   +2 \xi \df \xi  + 2 \chi_\xi \df {\chi_\xi} +
\partial^\mu \df {H^\mu} \Bigr) \nonumber \\
{\cal N}^{(l)}_\lambda & = & \Tr \intd g^{2l} f_\lambda(\xi)\Bigl(
\lambda^\alpha \df{\lambda^\alpha} + \lambdabar^\alphadot
\df{\lambdabar^\alphadot}
 -Y_\lambda^\alpha \df{Y_\lambda^\alpha} - Y_\lambdabar^\alphadot
\df{Y_\lambdabar^\alphadot} \Bigr) \nonumber \\
{\cal N}^{(l)}_{\lambda A} & = &
\Tr \intd i g^ {2(l+1)}f_{\lambda A}(\xi)
\Bigl( A_\mu ( \sigma^
\mu \chibar )^ \alpha \df {{\lambda}^ { \alpha}} + Y_{\lambda}^ { \alpha} \sigma^
\mu_{\alpha
\alphadot} \chibar^ \alphadot
\df {\rho_\mu} - \mbox{c.c.}\Bigr) \nonumber
\end{eqnarray}
\item
The symmetric differential operator of the Faddeev--Popov ghost:
\begin{eqnarray}
{\cal N}^{(l)}_c \Ga+ \Delta^{(l)}_{Y, c} &= &
\bigl(\brs^{r_\eta}_\Ga + r^{(1)} \delta {\cal S}\bigr)\Tr \intd g^{2l} 
(- Y_c c )
\end{eqnarray}
and 
\begin{equation}
{\cal N}^{(l)}_c = \Tr \intd g^{2l} \Bigl( c \df c - Y_c \df
{Y_c}\Bigr)
\end{equation}
\item The symmetric differential operators of the matter fields:
\begin{eqnarray}
\label{ctvarphi} 
{\cal N}^ {(l)}_\varphi \Ga + \Delta^ {(l)}_{Y,\varphi}
 & \equiv & \tilde \brs_\Ga  \intd g^{2l} f^ {(l)}_{\varphi}(\tilde \xi)
\bigl(Y_{\varphi_L} \varphi_L  +
Y_{\phibar_L} \phibar_L +  (_{L\to R})  \bigr)  \\
\label{ctpsi} 
{\cal N}^ {(l)}_\psi \Ga + \Delta^ {(l)}_{Y,\psi}
 & \equiv & \tilde \brs_\Ga   \intd g^{2l} f^ {(l)}_{\psi}(\tilde \xi)
\bigl(\psi_L Y_{\psi_L}    +
\psibar_L Y_{\psibar_L}  + (_{L\to R})  \bigr)   \nonumber \\
{\cal N}^ {(l)}_{\psi\varphi} \Ga + \Delta^
{(l)}_{Y,\psi\varphi} & \equiv &
\tilde \brs_\Ga 
 \intd \sqrt 2  f^ {(l)}_{\psi\varphi}(\tilde \xi)
\bigl(\chi^ {E^ {(2l)}} \varphi_L Y_{\psi_L}    \nonumber \\
& & \phantom{( s_{\Ga} + r^ {(1)} \delta {\cal S})
 \intd \sqrt 2} + {\chibar^  {E^ {(2l)}}}
\phibar_L Y_{\psibar_L} + (_{L\to R})\bigr)  \nonumber 
\label{Nende}
\end{eqnarray}
Explicit expressions are immediately obtained by evaluating the
variation. They can be also
found in \cite{KRST01}.
\end{itemize}

\end{appendix}


\end{document}